\documentclass[preprint,
times,
twoside]{elsarticle}

\usepackage{booktabs}
\usepackage{amsmath,amssymb}
\usepackage{amsthm}
\usepackage{mathrsfs}
\usepackage{algorithm2e}
\input epsf
\usepackage{epsfig}
\usepackage{verbatim}
\usepackage{a4wide,examplep}
\usepackage{multirow}
\usepackage{graphicx}

\newtheorem{theorem}{Theorem}
\newtheorem{lemma}[theorem]{Lemma}
\newtheorem{proposition}{Proposition}

\newtheorem{remark}{Remark}
\newtheorem{corollary}{Corollary}
\def\RR{\mathbb R}
\def\xxi{\boldsymbol{\xi}}
\def\zzeta{\boldsymbol{\zeta}}
\def\bY{\boldsymbol{Y}}
\def\T{\top}
\def\Ex{\mathbf E}
\def\KL{\mathop{\mathcal K}}
\def\1{\mathbf 1}

\def\sign{\mathop{\text{sign}}}
\def\xxi{{\boldsymbol{\xi}}}
\def\zzeta{{\boldsymbol{\zeta}}}
\def\llambda{{\boldsymbol{\lambda}}}
\def\llambdastar{{\llambda^{*}}}

\def\bbX{{\mathbb{X}}}
\def\ttA{\texttt{A}}

\def\opi{\bar\omega}

\def\Pb{{\mathbf P}}

\def\bY{{\mathbf Y}}
\def\bL{{\boldsymbol L}}

\def\bW{{\boldsymbol W}}
\def\bB{{\boldsymbol B}}

\def\bv{{\boldsymbol v}}

\def\boldh{{\boldsymbol h}}

\def\bx{{\boldsymbol x}}

\def\bX{{\boldsymbol X}}

\def\RR{{\mathbb R}}

\def\T{\top}
\def \1{{\rm 1}\mskip -4,5mu{\rm l} }

\def\tr{{\rm Tr}}
\def\KL{\mathcal K}
\def\sign{\mathop{\rm sgn}}

\def\leqserre{\!\!\!\!&\le&\!\!\!\!}
\def\eqserre{\!\!\!\!\!&=&\!\!\!\!}

\def\mcZ{\mathcal Z}

\def\mcFL{\mathcal F_\Lambda}

\def\Ex{\mathbf E}
\def\Cov{\mathbf{Cov}}

\def\mcPL1{\mathcal P_1(\mcFL)}
\def\flambda{f_{\lambda}}
\def\fllambda{f_{\llambda}}
\def\le{\leqslant}
\def\ge{\geqslant}

\def\lj{\lambda_{\!j}}

\def\hat{\widehat}

\begin{document}

\title{Sparse Regression Learning by Aggregation and Langevin Monte-Carlo}

\author[UMLV]{A.S.~Dalalyan}
\ead{dalalyan@imagine.enpc.fr}
\author[LPMA]{A.B.~Tsybakov}
\ead{alexandre.tsybakov@ensae.fr}

\address[UMLV]{IMAGINE, LIGM, Universit\'e Paris Est, Ecole des Ponts ParisTech, FRANCE\\}
\address[LPMA]{CREST and LPMA, Universit\'e Paris 6,FRANCE}

\begin{abstract}
We consider the problem of regression learning for deterministic
design and independent random errors. We start by proving a sharp
PAC-Bayesian type bound for the exponentially weighted aggregate
(EWA) under the expected squared empirical loss. For a broad class
of noise distributions the presented bound is valid whenever the
temperature parameter $\beta$ of the EWA is larger than or equal
to $4\sigma^2$, where $\sigma^2$ is the noise variance. A
remarkable feature of this result is that it is valid even for
unbounded regression functions and the choice of the temperature
parameter depends exclusively on the noise level.

Next, we apply this general bound to the problem of aggregating the
elements of a finite-dimensional linear space spanned by a
dictionary of functions $\phi_1,\ldots,\phi_M$. We allow $M$ to be
much larger than the sample size $n$ but we assume that the true
regression function can be well approximated by a sparse linear
combination of functions $\phi_j$. Under this sparsity scenario, we
propose an EWA with a heavy tailed prior and we show that it
satisfies a sparsity oracle inequality with leading constant one.

Finally, we propose several Langevin Monte-Carlo algorithms to
approximately compute such an EWA when the number $M$ of
aggregated functions can be large. We discuss in some detail the
convergence of these algorithms and present numerical experiments
that confirm our theoretical findings.
\end{abstract}


\journal{JCSS}

\maketitle

\section{Introduction}

In recent years a great deal of attention has been devoted to
learning in high-dimensional models under the sparsity scenario.
This typically assumes that, in addition to the sample, we have a
finite dictionary of very large cardinality such that a small set
of its elements provides a nearly complete description of the
underlying model. Here, the words ``large'' and ``small'' are
understood in comparison with the sample size. Sparse learning
methods have been successfully applied in bioinformatics,
financial engineering, image processing, etc. (see, e.g., the
survey in \cite{Y07}).

A popular model in this context is linear regression. We observe $n$
pairs $(\bX_1,Y_1),\ldots,(\bX_n,Y_n)$, where each $\bX_i$ -- called
the predictor -- belongs to $\RR^M$ and $Y_i$ -- called the response
-- is scalar and satisfies $Y_i=\bX_i^\T \llambda_0 + \xi_i$ with
some zero-mean noise $\xi_i$. The goal is to develop inference on
the unknown vector $\llambda_0\in\RR^M$.

In many applications of linear regression the dimension of $\bX_i$
is much larger than the sample size, i.e.,\ $M\gg n$. It is
well-known that in this case classical procedures, such as the
least squares estimator, do not work. One of the most compelling
ways for dealing with the situation where $M\gg n$ is to suppose
that the sparsity assumption is fulfilled, i.e., that
$\llambda_0$ has only few coordinates different from $0$. This
assumption is helpful at least for two reasons: The model becomes
easier to interpret and the consistent estimation of $\llambda_0$
becomes possible if the number of non-zero coordinates is small
enough.

During the last decade several learning methods exploiting the
sparsity assumption have been discussed in the literature. The
$\ell_1$-penalized least squares (Lasso) is by far the most studied
one and its statistical properties are now well understood (cf.,
e.g., \cite{BRT08,BTW06,BTW07a,BTW07b,MeinBuhl06,vdGeer08,ZH08} and
the references cited therein). The Lasso is particularly attractive
by its low computational cost. For instance, one can use the LARS
algorithm \cite{LARS}, which is quite popular. Other procedures
based on closely related ideas include the Elastic Net
\cite{HasZou05}, the Dantzig selector \cite{CT07} and the least
squares with entropy penalization \cite{Kol08}. However, one
important limitation of these procedures is that they are provably
consistent under rather restrictive assumptions on the Gram matrix
associated to the predictors, such as the mutual coherence
assumption \cite{DET06}, the uniform uncertainty principle
\cite{CT06}, the irrepresentable \cite{ZhaoYu06} or the restricted
eigenvalue \cite{BRT08} conditions. This is somewhat unsatisfactory,
since it is known that, at least in theory, there exist estimators
attaining optimal accuracy of prediction under almost no assumption
on the Gram matrix. This is, in particular, the case for the
$\ell_0$-penalized least squares estimator \cite[Thm.~3.1]{BTW07a}.
However, the computation of this estimator is an NP-hard problem. We
finally mention the paper \cite{Wipf}, which brings to attention the
fact that the empirical Bayes estimator in Gaussian regression with
Gaussian prior can effectively recover the sparsity pattern. This
method is realized in \cite{Wipf} via the EM algorithm. However, its
theoretical properties are not explored, and it is not clear what
are the limits of application of the method beyond the considered
set of numerical examples.

In~\cite{DT07, DT08} we proposed another approach to learning under
the sparsity scenario, which consists in using an exponentially
weighted aggregate (EWA) with a properly chosen sparsity-favoring
prior. There exists an extensive literature on EWA. Some recent
results focusing on the statistical properties can be found in
\cite{Al08,Aud08,Cat07,JRT08,LB06,Yang04}. Application of EWA
to the single-index regression and Gaussian graphical models has been
developed in \cite{GL07} and \cite{GHV09}, respectively. Procedures with
exponential weighting received much attention in the literature on
the on-line learning, see \cite{CBCG04,HKW98,Vovk90}, the monograph
\cite{CBL06} and the references cited therein.

The main message of \cite{DT07,DT08} is that the EWA with a properly
chosen prior is able to deal with the sparsity issue. In particular,
\cite{DT07,DT08} prove that such an EWA satisfies a sparsity oracle
inequality (SOI), which is more powerful than the best known SOI for
other common procedures of sparse recovery. An important point is
that almost no assumption on the Gram matrix is required. In the
present work we extend this analysis in two directions. First, we
prove a sharp PAC-Bayesian bound for a large class of noise
distributions, which is valid for the temperature parameter
depending only on the noise distribution. We impose no restriction
on the values of the regression function. This result is presented
in Section~\ref{S2}. The consequences in the context of linear
regression under sparsity assumption are discussed in
Section~\ref{S3}.

The second problem that we analyze here is the computation of EWA
with the sparsity prior. Since we want to deal with large dimensions
$M$, computation of integrals over $\RR^M$ in the definition of this
estimator can be a hard problem. Therefore, we suggest an
approximation based on Langevin Monte-Carlo (LMC). This is described
in detail in Section~\ref{S4}. Section~\ref{S5} contains numerical
experiments that confirm fast convergence properties of the LMC and
demonstrate a nice performance of the resulting estimators.

\section{PAC-Bayesian type oracle inequality}\label{S2}

 Throughout this section, as well as in Section \ref{S3}, we assume that
we are given the data $(Z_i,Y_i),\,i=1,\ldots,n$, generated by the
non-parametric regression model
\begin{equation}\label{model}
Y_i=f(Z_i)+\xi_i,\qquad i=1,\ldots,n,
\end{equation}
with deterministic design $Z_1,\ldots,Z_n$ and random errors
$\xi_i$. We use the vector notation $\bY=\boldsymbol{f}+\xxi$, where
$\xxi= (\xi_1,\dots,\xi_n)^\T$ and the function $f(\cdot)$ is
identified with the vector $\boldsymbol
f=(f(Z_1),\ldots,f(Z_n))^\T$. The space $\mathcal Z$ containing the
design points $Z_i$ can be arbitrary and $f$ is a mapping from
$\mathcal Z$ to $\RR$. For each function $h:\mathcal Z\to \RR$, we
denote by $\|h\|_n$ the empirical norm $\big(\frac1n\sum_{i=1}^n
h(Z_i)^2\big)^{1/2}$. Along with these notation, we will denote by
$\|\bv\|_p$ the $\ell_p$-norm of a vector
$\bv=(v_1,\dots,v_n)\in\RR^n$, that is $\|\bv\|_p^p=\sum_{i=1}^n
|v_i|^p$, $1\le p<\infty$, $\|\bv\|_\infty=\max_{i} |v_i|$ and
$\|\bv\|_0$ is the number of nonzero entries of $\bv$. With this
notation, $\|\boldsymbol f\|_2^2=n\|f\|_n^2$.

Assume that we are given a collection $\{f_\lambda:\lambda\in\Lambda\}$ of
functions $f_\lambda:\mathcal Z\to\RR$ that will serve as building
blocks for the learning procedure.The set $\Lambda$ is assumed to be
equipped with a $\sigma$-algebra and the mappings $\lambda\mapsto
f_\lambda(z)$ are assumed to be measurable with respect to this
$\sigma$-algebra for all $z\in \mathcal Z$. Let $\pi$ be a
probability measure on $\Lambda$, called the prior, and let $\beta$
be a positive real number, called the temperature parameter. We
define the EWA by
$$
\hat f_n(z)=\int_\Lambda f_\lambda(z)\,\hat\pi_{n,\beta}(d\lambda),
$$
where $\hat\pi_{n,\beta}$ is the (posterior) probability distribution
$$
\hat\pi_{n,\beta}(d\lambda)\propto \exp\big\{-\beta^{-1}\|\bY-
\boldsymbol{f}_\lambda\|_2^2\big\}\,\pi(d\lambda),
$$
and $\boldsymbol
f_{\lambda}=(f_{\lambda}(Z_1),\ldots,f_{\lambda}(Z_n))^\T$. We
denote by $L$ the smallest positive number, which may be equal to
$+\infty$, such that
\begin{equation}\label{L}
(\lambda,\lambda')\in\Lambda^2\quad \Longrightarrow\quad \max_i |f_\lambda(Z_i)-f_{\lambda'}(Z_i)|\le L
\end{equation}
In the sequel, we use the convention $\frac{+\infty}{+\infty}=0$
and, for any function $v:\RR\to\RR$, we denote by $\|v\|_\infty$ its
$L_\infty(\RR)$-norm.

In order to get meaningful statistical results on the accuracy of
the EWA, some conditions on the noise are imposed. In addition to
the standard assumptions that the noise vector
$\xxi=(\xi_1,\ldots,\xi_n)^\T$ has zero mean and independent
identically distributed (iid) coordinates, we require the following
assumption on the distribution of $\xi_1$.

\medskip
\noindent \textbf{Assumption N.} For any $\gamma>0$ small enough,
there exist a probability space and two random variables $\xi$ and
$\zeta$ defined on this probability space such that\\[-19pt]
\begin{itemize}
\item[i)] $\xi$ has the same distribution  as the regression errors $\xi_i$, \\[-19pt]
\item[ii)] $\xi+\zeta$ has the same distribution as $(1+\gamma)\xi$
and the conditional expectation satisfies $\Ex[\zeta\,|\,\xi]=0$,\\[-19pt]
\item[iii)] there exist $t_0\in(0,\infty]$ and a bounded Borel
function $v:\RR\to\RR_+$ such that
$$
\varlimsup_{\gamma\to 0} \ \sup_{(t,a)\in[-t_0,t_0]\times {\rm supp}
(\xi)} \frac{\log \Ex[e^{t\zeta}\,|\,\xi=a]}{t^2\gamma v(a)}\le 1,
$$
where ${\rm supp} (\xi)$ is the support of the distribution of
$\xi$.
\end{itemize}
Many symmetric distributions used in applications satisfy Assumption
N with functions $v$ such that $\|v\|_\infty$ is a multiple of the
variance of the noise $\xi_i$. This follows from
Remarks~\ref{rem:1}-\ref{rem:6} given at the end of this section and
their combinations.

%%%%%%%%%%%%%%%%%%%%%%%%%%%%%%%%%%%%%%%%%%%%%%%%%%%%%%%%%%%%%%%%%%%%%%%%%%%%
%%%%%%%%%%%%%%%%%%%%%%%%%%%%%%%%%%%%%%%%%%%%%%%%%%%%%%%%%%%%%%%%%%%%%%%%%%%%
\begin{theorem}\label{Thm1}
Let Assumption N be satisfied with some function $v$ and
let (\ref{L}) hold. Then for any prior $\pi$, any probability
measure $p$ on $\Lambda$ and any $\beta\ge
\max(4\|v\|_\infty,2L/t_0)$ we have
$$
\Ex[\|\hat f_n-f\|_n^2]\le
\int_\Lambda \|f-f_\lambda\|_n^2\,p(d\lambda)+\frac{\beta \KL(p,\pi)}n,
$$
where $\KL(\cdot\,,\cdot\;\!)$ stands for the Kullback-Leibler
divergence.
\end{theorem}

Prior to presenting the proof, let us note that Theorem~\ref{Thm1}
is in the spirit of \cite[Theorems~1,2]{DT08}, but is better in
several aspects. First, the main assumption ensuring the validity of
the oracle inequality involves the distribution of the noise alone,
while \cite[Theorem~2]{DT08} relies on an assumption (denoted by
\textbf{(C)} in \cite{DT08}) that ties together the distributional
properties of the noise and the nature of the dictionary
$\{\flambda\}$. A second advantage is that Assumption N is
independent of the sample size $n$ and, consequently, suggests a
choice of the parameter $\beta$ that does not change with the sample
size. Theorem~1 of \cite{DT08} also has these advantages but it is
valid only for a very restricted class of noise distributions,
essentially for the Gaussian and uniform noise. As we shall see
later in this section, Theorem~\ref{Thm1} leads to a choice of the
tuning parameter $\beta$, which is very simple and guarantees the
validity of a strong oracle inequality for a large class of noise
distributions.

\begin{proof}[Proof of Theorem~\ref{Thm1}]
It suffices to prove the theorem for $p$  such that $\int_\Lambda
\|f_\lambda-f\|_n^2\,p(d\lambda)<\infty$ and $p\ll \pi$ (implying
$\KL(p,\pi)<\infty$), since otherwise the result is trivial.

We first assume that $\beta>4\|v\|_\infty$ and that $L<\infty$.
Let $\gamma>0$ be a small number. Let now
$(\xi_1,\zeta_1),\ldots,(\xi_n,\zeta_n)$ be a sequence of iid
pairs of random variables defined on a common probability space
such that $(\xi_i,\zeta_i)$ satisfy conditions i)-iii) of
Assumption N for any $i$. The existence of these random variables
is ensured by Assumption N. We use here the same notation $\xi_i$
as in model (\ref{model}), since it causes no ambiguity.

Set $\boldh_\lambda=f_\lambda-f$, $\hat h=\hat f_n-f$,
$\zzeta=(\zeta_1,\dots,\zeta_n)^\T$,
$U(\boldh,\boldh')=\|\boldh\|_2^2+2 {\boldh\!}^\T\boldh'$ and $\Delta
U(\boldh,\boldh',\boldh'')=(\|\boldh\|_2^2-\|\boldh'\|_2^2)+2(\boldh-
\boldh')^\T\boldh''$ for any
pair $\boldh,\boldh',\boldh''\in\RR^n$. With this notation we have
\begin{equation*}
\Ex[\|\hat f_n-f\|_n^2]=\Ex[\|\hat h\|_n^2]=\Ex\Big[\|\hat h\|_n^2+
\frac2{n\gamma}\hat\boldh^\T\!\zzeta\Big].
\end{equation*}
Therefore, $\Ex[\|\hat f_n-f\|_n^2]=S+S_1$, where
\begin{eqnarray*}
S\!\!\!\!&=&\!\!\!-\frac\beta{n\gamma}\Ex\Big[\log\int_\Lambda\exp\!\Big(\!-\frac{\gamma U(\boldh_\lambda,\gamma^{-1}\zzeta)}{\beta}\Big)\,\widehat\pi_{n,\beta}(d\lambda)\Big],\\
S_1\!\!\!\!\!&=&\!\!\!\frac\beta{n\gamma}\Ex\Big[\log\!\!\int_\Lambda\exp\!\Big(\!-\frac{\gamma\Delta U(\boldh_\lambda,\hat\boldh,\gamma^{-1}\zzeta)}{\beta}\Big)\,\widehat\pi_{n,\beta}(d\lambda)\Big].
\end{eqnarray*}
We first bound the term $S$. To this end, note
that
$$
\widehat\pi_{n,\beta}(d\lambda)=
\frac{\exp\{-\beta^{-1}U(\boldh_\lambda,\xxi)\}}{\int_\Lambda \exp\{-\beta^{-1}U(\boldh_w,\xxi)\}\pi(dw)}\;\pi(d\lambda)
$$
and, therefore,
\begin{eqnarray*}
S\!\!\!&=&\!\!\!\frac\beta{n\gamma}\Ex\Big[\!\log\int_\Lambda \exp\big\{\!\!-\hbox{$\frac1\beta$}U(\boldh_\lambda,\xxi)\big\}\pi(d\lambda)\Big]-\frac\beta{n\gamma}\Ex\Big[\!\log\!\int_\Lambda\! \exp\big\{-\hbox{$\frac{1+\gamma}{\beta}$}U\big(\boldh_\lambda,\hbox{$\frac{\xxi+\zzeta}{1+\gamma}$}\big)\big\}\pi(d\lambda)\Big].
\end{eqnarray*}
By part ii) of Assumption N and the independence of vectors
$(\xi_i,\zeta_i)$ for different values of $i$, the probability
distribution of the vector $(\xxi+\zzeta)/(1+\gamma)$ coincides with
that of $\xxi$. Therefore, $(\xxi+\zzeta)/(1+\gamma)$ may be
replaced by $\xxi$ inside the second expectation. Now, using the
H\"older inequality, we get
\begin{eqnarray*}
S\leqserre-\frac\beta{n(1+\gamma)} \Ex\Big[\log\int_\Lambda e^{-(1+\gamma)\beta^{-1}U(\boldh_\lambda,\xxi)}\pi(d\lambda)\Big].
\end{eqnarray*}
Next, by a convex duality argument \cite[p.\ 160]{Cat04}, we find
$$
S\le \int_\Lambda \| h_\lambda\|_n^2\,p(d\lambda)+\frac{\beta \KL(p,\pi)}{n(1+\gamma)}\ .
$$
Let us now bound the term $S_1$. According to part iii) of
Assumption N, there  exists $\gamma_0>0$ such that
$\forall\gamma\le\gamma_0$,
$$
\sup_{|t|\le t_0}
\frac{\log \Ex[e^{t\zeta}|\xi=a]}{t^2\gamma}\le v(a)(1+o_\gamma(1)),\quad \forall\,a\in\RR.
$$
In what follows we assume that $\gamma\le\gamma_0$. Since for every
$i$, $|2\beta^{-1}(h_\lambda(Z_i)-\hat h(Z_i))|\le 2\beta^{-1}L\le
t_0$, using Jensen's inequality we get
\begin{eqnarray*}
S_1\leqserre \frac\beta{n\gamma}\Ex\Big[\log\!\int_\Lambda \exp\Big\{-\frac{n\gamma}{\beta}(\|h_\lambda\|_n^2-
\|\hat h\|_n^2)\Big\}\,\theta_\lambda\,\Ex\!\Big(\!\exp\!\Big\{\!\sum_{i=1}^n 2\beta^{-1}(h_\lambda(Z_i)-\hat h(Z_i))\zeta_i\!\Big\}\big|{\xxi}\Big)\pi(d\lambda)\Big]\\
\leqserre \frac\beta{n\gamma}\Ex\Big[\log\!\int_\Lambda \exp\Big\{-\frac{n\gamma}{\beta}(\|h_\lambda\|_n^2-
\|\hat h\|_n^2)\Big\}\,\theta_\lambda\,\exp\Big\{\frac{4n\|v\|_\infty\gamma}{\beta^2}\|h_\lambda-\hat h\|^2_n(1\!+o_\gamma(1))\Big\}\, \pi(d\lambda)\Big].
\end{eqnarray*}
For $\gamma$ small enough ($\gamma\le\tilde\gamma_0$), this entails that up to a
positive multiplicative constant, the term $S_1$ is
bounded by the expression $\Ex\big[\log\int_\Lambda \exp\big(-\frac{n\gamma{V}(h_\lambda,\hat h)}{\beta^2} \big)
\theta_\lambda\pi(d\lambda)\big]$, where
$$
V(h_\lambda,\hat h)\!=\beta(\| h_\lambda\|_n^2-\|\hat h\|_n^2)\!+\frac{(\beta+4\|v\|_\infty)}{2}\|h_\lambda\!-\hat h\|^2_n.
$$
Using \cite[Lemma 3]{DT07} and Jensen's inequality we obtain
$S_1\le 0$ for any $\gamma\le (\beta-4\|v\|_\infty)/4nL$. Thus,
we proved that
$$
\Ex[\|\hat h\|_n^2]\le \int_\Lambda \|h_\lambda\|_n^2\,p\,(d\lambda)+\frac{\beta\, \KL(p,\pi)}{n(1+\gamma)}
$$
for any $\gamma\le\tilde\gamma_0\wedge (\beta-4\|v\|_\infty)/4nL$.
Letting $\gamma$ tend to zero, we obtain
$$
\Ex[\|\hat h\|_n^2]\le \int_\Lambda \|h_\lambda\|_n^2p(d\lambda)+\frac{\beta\, \KL(p,\pi)}n
$$
for any $\beta
>\max(4\|v\|_\infty,2L/t_0)$. Fatou's lemma allows us to extend this
inequality to the case $\beta=\max(4\|v\|_\infty,2L/t_0)$.

To cover the case $L=+\infty,t_0=+\infty$, we fix some
$L_0\in(0,\infty)$ and apply the obtained inequality to the
truncated prior $\pi^{L'}(d\lambda)\propto
\1_{\Lambda_{L'}}(\lambda)\pi(d\lambda)$, where
$L'\in(L_0,\infty)$ and $\Lambda_{L'}=\{\lambda\in\Lambda: \max_i
|f_\lambda(Z_i)|\le L'\}$. We obtain that for any measure
$p\ll\pi$ supported by $\Lambda_{L_0}$,
\begin{eqnarray*}
\Ex[\|\hat h^{L'}\|_n^2]\leqserre \int_\Lambda \|h_\lambda\|_n^2\,p(d\lambda)+\frac{\beta \KL(p,\pi^{L'})}n\\
\leqserre \int_\Lambda \|h_\lambda\|_n^2\,p(d\lambda)+\frac{\beta \KL(p,\pi)}n.
\end{eqnarray*}
One easily checks that $\hat h^{L'}$ tends a.s.\ to $\hat h$ and
that the random variable $\sup_{L'>L_0} \|\hat
h^{L'}\|_n^2\1(\max_i |\xi_i|\le C)$ is integrable for any fixed
$C$. Therefore, by Lebesgue's dominated convergence theorem we get
\begin{eqnarray*}
\Ex[\|\hat h\|_n^2\1(\max_i |\xi_i|\le C)]\leqserre \int_\Lambda \|h_\lambda\|_n^2\,p(d\lambda)+\frac{\beta \KL(p,\pi)}n.
\end{eqnarray*}
Letting $C$ tend to infinity and using Lebesgue's monotone
convergence theorem we obtain the desired inequality for any
probability measure $p$ which is absolutely continuous w.r.t.\
$\pi$ and is supported by $\Lambda_{L_0}$ for some $L_0>0$. If
$p(\Lambda_{L_0})<1$ for any $L_0>0$, one can
replace $p$ by its truncated version $p^{L'}$ and use Lebesgue's
monotone convergence theorem to get the desired result.
\end{proof}

The following remarks provide examples of noise distributions, for
which Assumption N is satisfied. Proofs of these remarks are given
in the Appendix.

\begin{remark}[Gaussian noise]\label{rem:1} If $\xi_1$ is drawn according to the Gaussian distribution $\mathcal N(0,\sigma^2)$,
then for any $\gamma>0$ one can choose $\zeta$ independently of
$\xi$ according to the Gaussian distribution $\mathcal
N(0,(2\gamma+\gamma^2)\sigma^2)$. This results in $v(a)\equiv
\sigma^2$ and, as a consequence, Theorem~\ref{Thm1} holds for any
$\beta\ge 4\sigma^2$. Note that this reduces to the Leung and
Barron's \cite{LB06} result if the prior $\pi$ is discrete.
\end{remark}

\begin{remark}[Rademacher noise]\label{rem:2} If $\xi_1$ is drawn according to the Rademacher distribution, i.e.\
$\Pb(\xi_1=\pm\sigma)=1/2$, then for any $\gamma>0$ one can define
$\zeta$ as follows:
$$
\zeta=(1+\gamma)\sigma\sign[\sigma^{-1}\xi-(1+\gamma)U]-\xi,
$$
where $U$ is distributed uniformly in $[-1,1]$ and is independent of
$\xi$. This results in $v(a)\equiv \sigma^2$ and, as a consequence,
Theorem~\ref{Thm1} holds for any $\beta\ge 4\sigma^2=4\Ex[\xi_1^2]$.
\end{remark}

\begin{remark}[Stability by convolution] Assume that $\xi_1$ and $\xi_1'$ are two independent random variables.
If $\xi_1$ and $\xi_1'$ satisfy Assumption N with $t_0=\infty$ and
with functions $v(a)$ and $v'(a)$, then any linear combination
$\alpha\xi_1+\alpha'\xi_1'$ satisfies Assumption N with $t_0=\infty$
and the $v$-function $\alpha^2 v(a)+(\alpha')^2v'(a)$.
\end{remark}

\begin{remark}[Uniform distribution] The claim of preceding remark
can be generalized to linear combinations of a countable set of
random variables, provided that the series converges in the mean
squared sense. In particular, if $\xi_1$ is drawn according to the
symmetric uniform distribution with variance $\sigma^2$, then
Assumption N is fulfilled with $t_0=\infty$ and $v(a)\equiv
\sigma^2$. This can be proved using the fact that $\xi_1$ has the
same distribution as $\sigma\sum_{i=1}^\infty 2^{-i}\eta_i$, where
$\eta_i$ are iid Rademacher random variables. Thus, in this case the
inequality of Theorem~\ref{Thm1} is true for any $\beta\ge
4\sigma^2$.
\end{remark}

\begin{remark}[Laplace noise] If $\xi_1$ is drawn according to the Laplace distribution with variance $\sigma^2$,
then for any $\gamma>0$ one can choose $\zeta$ independently of
$\xi$ according to the distribution associated to the
characteristic function
$$
\varphi(t)=\frac1{(1+\gamma)^2}\Big(1+\frac{2\gamma+\gamma^2}{1+(1+\gamma)^2(\sigma t)^2/2}\Big).
$$
One can observe that the distribution of $\zeta$ is a mixture of the
Dirac distribution at zero and the Laplace distribution with
variance $(1+\gamma)^2\sigma^2$. This results in $v(a)\equiv
2\sigma^2/(2-\sigma^2t_0^2)$ and, as a consequence, by taking
$t_0=1/\sigma^2$, we get that Theorem~\ref{Thm1} holds for any
$\beta\ge \max(8\sigma^2,2L\sigma)$.
\end{remark}

\begin{remark}[Bounded symmetric noise]\label{rem:6}
Assume that the errors $\xi_i$ are symmetric and that $P(|\xi_i|\le
B)=1$ for some $B\in(0,\infty)$. Let $U\sim\mathcal U([-1,1])$ be a
random variable independent of $\xi$. Then,
$\zeta=(1+\gamma)|\xi|\sign[\sign(\xi)-(1+\gamma)U]-\xi$ satisfies
Assumption N with $v(a)=a^2$. Since $\|v\|_\infty \le B^2$, we
obtain that Theorem~\ref{Thm1} is valid for any $\beta\ge 4B^2$.
\end{remark}

Consider now the case of finite $\Lambda$. W.l.o.g. we suppose that
$\Lambda=\{1,\dots, M\}$, $\{f_\lambda,
\lambda\in\Lambda\}=\{f_1,\dots, f_M\}$ and we take the uniform
prior $\pi(\lambda=j)=1/M$. From Theorem~\ref{Thm1} we immediately
get the following sharp oracle inequality for model selection type
aggregation.
\begin{corollary}\label{cor1}
Let Assumption N be satisfied with some function $v$ and let
(\ref{L}) hold. Then for the uniform prior $\pi(\lambda=j)=1/M,
j=1,\dots,M$, and any $\beta\ge \max(4\|v\|_\infty,2L/t_0)$ we have
$$
\Ex[\|\hat f_n-f\|_n^2]\le \min_{j=1,\dots,M}
\|f_j-f\|_n^2+\frac{\beta \log M}{n}.
$$
\end{corollary}
This corollary can be compared with bounds for combining procedures
in the theory of prediction of deterministic sequences
\cite{v:90,lw94, cetal, kw99, CBCG04, CBL06}. With our notation, the
bounds proved in these works can be written is the form
\begin{equation}\label{vovk}
\frac1{n}\sum_{i=1}^n(Y_i-f^*(Z_i))^2 \le
C_1\min_{j=1,\dots,M}\frac1{n}\sum_{i=1}^n(Y_i-f_j(Z_i))^2
+\frac{C_2 \log M}{n}.
\end{equation}
Here $f_j(Z_i)$ is interpreted as the value of $Y_i$ predicted by
the $j$th procedure, $f^*(Z_i)$ as an aggregated forecast, and
$C_1\ge1$, $C_2>0$ are constants. Such inequalities are proved under
the assumption that $Y_i$'s are deterministic and uniformly bounded.
When $C_1=1$, applying (\ref{vovk}) to random uniformly bounded
$Y_i$'s from model (\ref{model}) with $\Ex(\xi_i)=0$ and taking
expectations can yield an oracle inequality similar to that of
Corollary~\ref{cor1}. However, the uniform boundedness of $Y_i$'s
supposes that not only the noise $\xi_i$ but also the functions $f$
and $f_j$ are uniformly bounded. Bounds on $f$ should be a priori
known for the construction of the aggregated rule $f^*$ in
(\ref{vovk}) but in practice they are not always available. Our
results are free of this drawback because they hold with no
assumption on $f$. We have no assumption on the dictionary
$\{f_1,\dots, f_M\}$ neither.

\section{Sparsity prior and SOI}\label{S3}

In this section we introduce the sparsity prior and present a
sparsity oracle inequality (SOI) derived from Theorem \ref{Thm1}.

In what follows we assume that $\Lambda\subset\RR^M$ for some
positive integer $M$. We will use boldface letters to denote vectors
and, in particular, the elements of $\Lambda$. For any square matrix
\textsf{A}, let $\tr(\textsf{A})$ denote the trace (sum of diagonal
entries) of \textsf{A}. Furthermore, we focus on the particular case
where $\mcFL$ is the image of a convex polytope in $\RR^M$ by a \textit{link function}
$g:\RR\to\RR$. More specifically, we assume that, for some $R\in(0,+\infty]$ and for a
finite number of measurable functions $\big\{\phi_j\big\}_{j=1,\ldots,M}$,
$$
\mcFL=\Bigg\{\fllambda(z)=g\bigg(\sum_{j=1}^M \lj \phi_j(z)\bigg),\
\forall z\in\mcZ \Big|\ \llambda\in \RR^M\text{ satisfies
}\|\llambda\|_1\le R\,\Bigg\},
$$%
where $\|\llambda\|_1=\sum_j |\lj|$ stands for the $\ell_1$-norm.
The link function $g$ is assumed twice continuously differentiable and known. Typical examples
of link function include the linear function $g(x)=x$, the exponential function
$g(x)=e^x$, the logistic function $g(x)=e^x/(e^x+1)$, the cumulative distribution function of the standard
Gaussian distribution, and so on.

If, in addition, $f\in\mcFL$, then model (\ref{model}) reduces to
that of single-index regression with known link function. In the particular case of $g(x)=x$,
this leads to the linear regression defined in the Introduction. Indeed, it
suffices to take
$$
\bX_i=(\phi_1(Z_i),\ldots,\phi_M(Z_i))^\T,\quad i=1,\ldots,n.
$$
This notation will be used in the rest of the paper along with the
assumption that $\bX_i$ are normalized so that all the diagonal
entries of matrix $\frac1n \sum_{i=1}^n \bX_i\bX_i^\T$ are equal to
one.

The family $\mcFL$ defined above satisfies inequality (\ref{L}) with $L=2R\|g'\|_\infty L_\phi$,
where $L_\phi=\max_{i,j} |\phi_j(Z_i)|$ and $\|g'\|_\infty$ is the maximum of the derivative of $g$
on the interval $[-RL_\phi,RL_\phi]$. Indeed, since $\Lambda$ is the $\ell_1$ ball of radius $R$ in
$\RR^M$ and $\phi_j$s are bounded by $L_\phi$, the real numbers $u_i=\llambda^\T \bX_i$ and
$u'_i={\llambda'}^\T \bX_i$ belong to the interval $[-RL_\phi,RL_\phi]$ for every $\llambda$ and
$\llambda'$ from $\Lambda$. Consequently, $|f_\llambda(Z_i)-f_{\llambda'}(Z_i)|=|g(u_i)-g(u_i')|=
\int _{u_i}^{u_i'}g'(s)\,ds$ is bounded by $\|g'\|_\infty |u_i-u_i'|$, the latter being smaller
than $2R\|g'\|_\infty L_\phi$.

We allow $M$ to be large, possibly much larger than the sample size
$n$. If $M\gg n$, we have in mind that the sparsity assumption
holds, i.e., there exists $\llambdastar\in\RR^M$ such that $f$ in
(\ref{model}) is close to $f_{\llambdastar}$ for some $\llambdastar$
having only a small number of non-zero entries. We handle this
situation via a suitable choice of prior $\pi$. Namely, we use a
modification of the sparsity prior proposed in \cite{DT07}. It
should be emphasized right away that we will take advantage of
sparsity for the purpose of prediction and not for data compression.
In fact, even if the underlying model is sparse, we do not claim
that our estimator is sparse as well, but we claim that it is quite
accurate under very mild assumptions. On the other hand, some
numerical experiments demonstrate the sparsity of our estimator and the fact
that it recovers correctly the true sparsity pattern in examples
where the (restrictive) assumptions mentioned in the Introduction
are satisfied (cf. Section \ref{S5}). However, our theoretical
results do not deal with this property.

%This point constitutes a conceptual difference between the procedure
%of aggregation described below and the procedures based on $\ell_1$
%techniques such as the LASSO, the Dantzig Selector and their
%analogues.

To specify the sparsity prior $\pi$ we need the Huber function
$\opi:\RR\to\RR$ defined by
$$
\opi(t)=
\begin{cases}
t^2, & \text{ if } |t|\le 1\\
2|t|-1, & \text{ otherwise. }
\end{cases}
$$
This function behaves very much like the absolute value of $t$, but
has the advantage of being differentiable at every point $t\in\RR$.
Let $\tau$ and $\alpha$ be positive numbers. We define the {\it
sparsity prior}
\begin{equation}\label{prior}
\pi(d\llambda)=\frac{\tau^{2M}}{C_{\alpha,\tau,R}}\Bigg\{\prod_{j=1}^M
\frac{e^{-\opi(\alpha\lj)}}{(\tau^2+\lj^2)^{2}}\Bigg\}
\1(\|\llambda\|_1\le R)\,d\llambda,
\end{equation}
where $C_{\alpha,\tau,R}$ is the normalizing constant.

Since the sparsity prior (\ref{prior}) looks somewhat complicated,
an heuristical explanation is in order. Let us assume that $R$ is
large and $\alpha$ is small so that the functions $e^{-\opi(\alpha
\lj)}$ and $\1(\|\llambda\|_1\le R)$ are approximately equal to one.
With this in mind, we can notice that $\pi$ is close to the
distribution of $\sqrt2\tau\bY$, where $\bY$ is a random vector
having iid\ coordinates drawn from Student's t-distribution with
three degrees of freedom. In the examples below we choose a very
small $\tau$, smaller than $1/n$. Therefore, most of the coordinates
of $\tau \bY$ are very close to zero. On the other hand, since
Student's t-distribution has heavy tails, a few coordinates of
$\tau\bY$ are quite far from zero.

\begin{figure}[t] %
\includegraphics[width=1\linewidth,height=120pt]{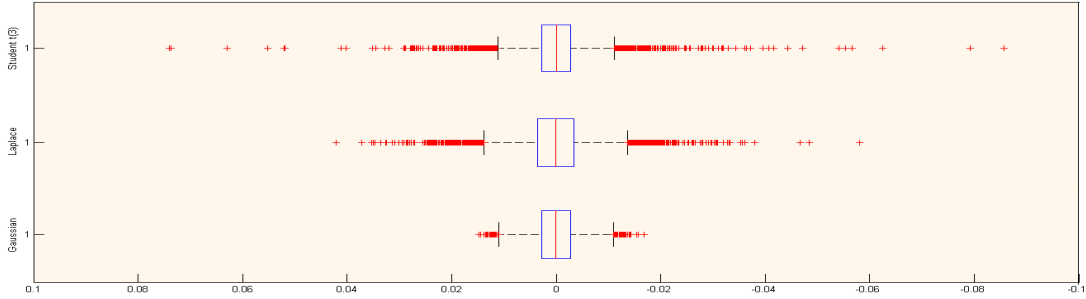}
\vglue-10pt \caption[]{The boxplots of a sample of size $10^4$ drawn
from the scaled Gaussian, Laplace and Student $t(3)$ distributions.
In all the three cases the location parameter is $0$ and the scale
parameter is $10^{-2}$.} \label{figSimu}
\end{figure}

These heuristics are illustrated by Figure~\ref{figSimu} presenting
the boxplots of one realization of a random vector in $\RR^{10.000}$
with iid\ coordinates drawn from the scaled Gaussian, Laplace
(double exponential) and Student $t(3)$ distributions. The scaling factor
is such that the probability densities of the simulated
distributions are equal to $100$ at the origin. The boxplot which is
most likely to represent a sparse vector corresponds to Student's
$t(3)$ distribution.

The relevance of heavy tailed priors for dealing with sparsity has
been emphasized by several authors (see \cite[Section 2.1]{Seeger08}
and references therein). However, most of this work focused on
logarithmically concave priors, such as the multivariate Laplace
distribution. Also in wavelet estimation on classes of ``sparse"
functions \cite{JS05} and \cite{Riv06} invoke quasi-Cauchy and
Pareto priors. Bayes estimators with heavy-tailed priors in sparse
Gaussian shift models are discussed in \cite{AGP07}.

The next theorem provides a SOI for the EWA with the sparsity prior
(\ref{prior}).
\begin{theorem}\label{ThmSparsity} Let Assumption N be satisfied
with some function $v$ and let (\ref{L}) hold. Take the
prior $\pi$ defined in (\ref{prior}) and $\beta\ge
\max(4\|v\|_\infty,2L/t_0)$. Assume that $R>2M\tau$ and
$\alpha\le 1/(4M\tau)$. Then for all $\llambdastar$ such that
$\|\llambdastar\|_1\le R-2M\tau$ we have
\begin{eqnarray}\nonumber
\Ex[\|\hat f_n-f\|_n^2]\leqserre
\|f_{\llambdastar}-f\|_n^2+\frac{4\beta}{n}\sum_{j=1}^M
\log\big(1+\hbox{$\frac{|\lj^*|}{\tau}$}\big)+\,\frac{2\beta(\alpha\|\llambdastar\|_1+1)}{n}+4eC_{g,f}\tau^2
M\label{ora}
\end{eqnarray}
with $C_{g,f}=1$ if $g(x)=x$ and $C_ {g,f}=\|g'\|_\infty^2+\|g''\|_\infty(\|g\|_\infty+\|f\|_\infty)$ for other
link functions $g$.
\end{theorem}
\begin{proof}
Let us define the probability measure $p_0$ by
\begin{equation}\label{eq:6}
\frac{dp_0}{d\llambda}(\llambda)\propto\left(\frac{d\pi}{d\llambda}(\llambda-\llambdastar)\right)\1_{B_1(2M\tau)}(\llambda-\llambdastar).
\end{equation}
Since $\|\llambdastar\|_1\le R-2M\tau$,
the condition $\llambda-\llambdastar\in B_1(2M\tau)$ implies that $\llambda\in B_1(R)$ and, therefore,
$p_0$ is absolutely continuous w.r.t.\ the sparsity prior $\pi$.  In view of Thm.~\ref{Thm1}, we have
$$
\Ex[\|\hat f_n-f\|^2_n]\le \int_\Lambda \|\fllambda-f\|_n^2\,p_0(d\llambda)+\frac{\beta\KL(p_0,\pi)}{n}.
$$
Since $\fllambda(Z_i)=g(\bX_i^\T\llambda)$ we have $\nabla_\llambda
[(\fllambda(Z_i)-f(Z_i))^2]=2g'(\bX_i^\T\llambda)(\fllambda(Z_i)-f(Z_i))\bX_i$
and
$$
\nabla^2_\llambda
[(\fllambda(Z_i)-f(Z_i))^2]=2\big\{g'(\bX_i^\T\llambda)^2+g''(\bX_i^\T\llambda)\big(g(\bX_i^\T\llambda)-f(Z_i)\big)\big\}\bX_i\bX_i^\T.
$$
One can remark that the factor of $\bX_i\bX_i^\T$ in the last
display is bounded by $C_{g,f}$. Therefore, in view of the Taylor
formula,
\begin{eqnarray*}
(\fllambda(Z_i)-f(Z_i))^2\leqserre (f_\llambdastar(Z_i)-f(Z_i))^2+2(f_\llambdastar(Z_i)-f(Z_i))g'(\bX_i^\T\llambdastar)\bX_i^\T(\llambda-\llambdastar)\\
&&\qquad+C_{g,f} [\bX_i^\T(\llambda-\llambdastar)]^2.
\end{eqnarray*}
By the symmetry of $p_0$ with respect to $\llambdastar$, the
integral $\int (\llambda-\llambdastar)p_0(d\llambda)$ vanishes.
Combining this with the fact that the diagonal entries of the matrix
$\frac1n\sum_i\bX_i\bX_i^\T$ are equal to one, we obtain
$$
\int_\Lambda \|\fllambda-f\|_n^2\,p_0(d\llambda)\le
\|f_{\llambda^*}-f\|_n^2+C_{g,f}\int_{\RR^M}\|\llambda-\llambda^*\|_2^2\,p_0(d\llambda).
$$
To complete the proof, we use the following technical result.
\begin{lemma}\label{lem:0}
For every integer $M$ larger than $1$, we have:
$$
\int_{\RR^M}\!\!\!
(\lambda_1-\lambda_1^*)^2p_0(d\llambda)\le 4\tau^2 e^{4M\alpha\tau},\quad
\KL(p_0,\pi)\le 2(\alpha\|\llambdastar\|_1+1)+4\sum_{j=1}^M \log(1+|\lj^*|/\tau).
$$
\end{lemma}
The proof of this lemma is postponed to the appendix. It is obvious that inequality (\ref{ora}) follows from Lemma 3, since
$\int_{\RR^M}\|\llambda-\llambda^*\|_2^2\,p_0(d\llambda)=M\int_{\RR^M}(\lambda_1-\lambda^*_1)^2\,p_0(d\llambda)$
and, under the assumptions of the theorem, $e^{4M\alpha\tau}\le e$.
\end{proof}

Theorem \ref{ThmSparsity} can be used to choose the tuning
parameters $\tau, \alpha,R$ when $M\gg n$. The idea is to choose
them such that both terms in the second line of (\ref{ora}) were of
the order $O(1/n)$. This can be achieved, for example, by taking
$\tau^2\sim (Mn)^{-1}$ and $R=O(M\tau)$. Then the term
$\frac{4\beta}{n}\sum_{j=1}^M \log (1+{|\lj^*|}/{\tau})$ becomes
dominating. It is important that the number $M^*$ of nonzero
summands
 in this term is equal to the number of nonzero coordinates of
$\llambdastar$. Therefore, for sparse vectors $\llambdastar$, this
term is rather small, namely of the order $M^*(\log M)/n$, which is
the same rate as achieved by other methods of sparse recovery, cf.
\cite{BTW06,CT07,BTW07a,BRT08}. An important difference compared
with these and other papers on $\ell_1$-based sparse recovery is
that in Theorem \ref{ThmSparsity}, we have no assumption on the
dictionary $\{\phi_1,\dots,\phi_M\}$.

Note that in the case of logistic regression the link function $g$,
as well as its first two derivatives, are bounded by one. Therefore,
since the logistic model is mainly used for estimating functions $f$
with values in $[0,1]$, Theorem~\ref{ThmSparsity} holds in this case
with $C_{g,f}\le 3$. Similarly, for the probit model (\textit{i.e.},
when the link function $g$ is the cdf of the standard Gaussian
distribution) and $f$ with values in $[0,1]$, one easily checks that
$C_{g,f}\le (\pi^{-1}+1)/2$.

\section{Computation of the EW-aggregate by the Langevin Monte-Carlo}\label{S4}

In this section we suggest Langevin Monte-Carlo (LMC) procedures to
approximately compute the EWA with the sparsity prior when $M\gg n$.

\subsection{Langevin Diffusion in continuous time} We start by describing a continuous-time
Markov process, called the Langevin diffusion, that will play the
key role in this section. Let $V:\RR^M\to \RR$ be a smooth function,
which in what follows will be referred to as potential. We will
assume that the gradient of $V$ is locally Lipschitz and is at most
of linear growth. This ensures that the stochastic differential
equation (SDE)
\begin{equation}\label{Langevin}
d\bL_t=\nabla V(\bL_t)\,dt+\sqrt{2}\,d\bW_t,\ \bL_0=\llambda_0,\ t\ge 0
\end{equation}
has a unique strong solution, called the Langevin diffusion. In the
last display, $\bW$ stands for an $M$-dimensional Brownian motion
and $\llambda_0$ is an arbitrary deterministic vector from $\RR^M$.
It is well known that the process $\{\bL_t\}_{t\ge 0}$ is a
homogeneous Markov process and a semimartingale, cf.\ \cite[Thm.
12.1]{RW}.

As a Markov process, $\bL$ may be transient, null recurrent or
positively recurrent. The latter case, which is the most important
for us, corresponds to the process satisfying the law of large
numbers and implies the existence of a stationary distribution. In
other terms, if $\bL$ is positively recurrent, there exists a
probability distribution $P_V$ on $\RR^M$ such that the process
$\bL$ is stationary provided that the initial condition $\llambda_0$
is drawn at random according $P_V$. A remarkable property of the
Langevin diffusion---making it very attractive for computing
high-dimensional integrals---is that its stationary distribution, if
exists, has the density
$$
p_V(\llambda)\propto e^{V(\llambda)},\qquad \llambda\in\RR^M,
$$
w.r.t.\ the Lebesgue measure \cite[Thm.\ 10.1]{Kent}. Furthermore,
some simple conditions on the potential $V$ ensure the positive
recurrence of $\bL$. The following proposition gives an example of
such a condition.

\begin{proposition} [\cite{RobStr03}, Thm 2.1]\label{PropDrift}
Assume that the function $V$ is bounded from above. If there is a
twice continuously differentiable function $D:\RR^M\to [1,\infty)$
and three positive  constants $a,b$ and $r$ such that
\begin{equation}\label{drift}
\nabla V(\llambda)^\T\nabla D(\llambda)+\Delta D(\llambda) \le -aD(\llambda)+b\1(\|\llambda\|_2\le r),
\end{equation}
for every $\llambda\in\RR^M$, then the Langevin diffusion $\bL$ defined
by (\ref{Langevin}) is $D$-geometrically ergodic, that is
$$
\Big|\Ex[h(\bL_t)|\bL_0=\llambda_0]-\int_{\RR^M} h(\llambda)
\,p_V(\llambda)d\llambda\Big|\le R_V D(\llambda_0)\rho^t_V
$$
for every function $h$ satisfying $\|h/D\|_\infty\le 1$ and for some constants $R_V>0$ and $\rho_V\in(0,1)$.
\end{proposition}

Function $D$ satisfying (\ref{drift}) is often referred to as
Lyapunov function and condition (\ref{drift}) is called the drift
condition towards the set $\{\llambda:\|\llambda\|_2\le r\}$. Recall
that the drift condition ensures geometrical mixing \cite[Theorem 16.1.5]{MeynTweedie}.
Specifically, for every function $h$ such that $\|h^2/D\|_\infty\le 1$ and for
every $t,s>0$,
$$
\big|\Cov_{\llambda_0}[h(\bL_t),h(\bL_s)]\big|\le R_VD(\llambda_0)\rho_V^{|t-s|}.
$$
Combining this with the result of Proposition~\ref{PropDrift} it is
not hard to check that if $\|h^2/D\|_\infty\le 1$, then
\begin{equation}\label{MSC}
\Ex_{\llambda_0}\bigg[\Big(\frac1T\! \int_0^T\!\! h(\bL_t)dt-
\!\int_{\RR^M}\!\! h(\llambda)p_V(\llambda)d\llambda\Big)^2\bigg]
\le \frac{C}{T},
\end{equation}
where $C$ is some positive constant depending only on $V$. Note also
that, in view of Proposition~\ref{PropDrift}, the squared bias term
in the bias-variance decomposition of the left hand side of
(\ref{MSC}) is of order $O(T^{-2})$. Thus, the main error term comes
from the stochastic part.

\subsection{Langevin diffusion associated to EWA}
In what follows, we focus on the particular case $g(x)=x$. Given
$(\bX_i,Y_i)$, $i=1,\ldots,n$, with $\bX_i\in\RR^M$ and $Y_i\in\RR$,
we want to compute the expression
\begin{equation}\label{6}
\hat\llambda=\frac{\int_{\RR^M} \llambda \exp\big\{-\beta^{-1}\|\bY-\mathbb X\llambda\|_2^2\big\}\pi(d\llambda)}
{\int_{\RR^M}\exp\big\{-\beta^{-1}\|\bY-\mathbb X\llambda\|_2^2\big\}\pi(d\llambda)},
\end{equation}
where $\mathbb X=(\bX_1,\ldots,\bX_n)^\T$. In what follows, we deal with the prior
$$
\pi(d\llambda)\propto\prod_{j=1}^M \frac{e^{-\opi(\alpha\lambda_j)}}{(\tau^2+\lambda_j^2)^2}
$$
assuming that $R=+\infty$. As proved in Sections~\ref{S2} and
\ref{S3}, this choice of the prior leads to sharp oracle
inequalities for a large class of noise distributions. An equivalent
form for writing (\ref{6}) is
$$\hat\llambda=\int_{\RR^M}\llambda p_V(\llambda)\,d\llambda, \quad
\text{where} \ p_V(\llambda)\propto e^{V(\llambda)}$$ with
\begin{equation}\label{V}
V(\llambda) = -\frac{\|\bY-\mathbb
X\llambda\|_2^2}{\beta}-\sum_{j=1}^M
\Big\{2\log(\tau^2+\lambda_j^2)+ \opi(\alpha\lambda_j)\Big\}.
\end{equation}

A simple algebra shows that $D(\llambda)=e^{\alpha\|\llambda\|_2}$
satisfies the drift condition (\ref{drift}). A nice property of this
Lyapunov function is the inequality $\|\llambda\|_\infty^2\le
\alpha^{-1}D(\llambda)$. It guarantees that (\ref{MSC}) is satisfied
for the functions $h(\llambda)=\lambda_i$.

Let us define the Langevin diffusion $\bL_t$ as solution of
(\ref{Langevin}) with the potential $V$ given in (\ref{V}) and the
initial condition $\bL_0=0$. In what follows we will consider only
this particular diffusion process. We define the average value
$$
\bar\bL_T=\frac1T\int_0^T \bL_t\,d t,\qquad T\ge 0.
$$
According to (\ref{MSC}) this average value converges as
$T\to\infty$ to the vector $\hat\llambda$ that we want to compute.
Clearly, it is much easier to compute $\bar\bL_T$ than
$\hat\llambda$. Indeed, $\hat\llambda$ involves integrals in $M$
dimensions, whereas $\bar\bL_T$  is a one-dimensional integral over
a finite interval. Of course, to compute such an integral one needs
to discretize the Langevin diffusion. This is done in the next
subsection.

\subsection{Discretization}\label{discr}

Since the sample paths of a diffusion process are H\"older continuous,
it is easy to show that the Riemann sum approximation
$$
\bar\bL_T^{R}=\frac1T\sum_{i=0}^{N-1}\bL_{T_i}\,(T_{i+1}-T_i),
$$
with $0=T_0<T_1<\ldots<T_N=T$ converges to $\bar\bL_T$ in mean
square when the sampling is sufficiently dense, that is when
$\max_{i}|T_{i+1}-T_i|$ is small. However, when simulating the
diffusion sample path in practice, it is impossible to follow
exactly the dynamics determined by (\ref{Langevin}). We need to
discretize the SDE in order to approximate the solution.

\begin{comment}
The theory of optimal discretization of continuous-time diffusions
is an active area of research, cf., e.g. \cite{someone}. Although it
suggests a variety of methods, there are no clear recommendations
for preferring one particular method to the others, even for a given
diffusion process with a perfectly specified drift function and
diffusion matrix. The main challenge in this context is to find the
best trade-off between the computational cost and the theoretical
accuracy of the discretization mechanism.
\end{comment}

%In this subsection we discuss different types of discretizations
%that can be useful for numerical approximation of exponentially
%weighted aggregates.
%
%
%
%\subsubsection{Constant step Euler discretization}
A natural discretization
for the SDE (\ref{Langevin}) is proposed by the Euler scheme with a constant step of
discretization $h>0$, defined as
\begin{equation}\label{CSE}
\bL_{k+1}^E=\bL_{k}^E+h\nabla V(\bL_k^E)+\sqrt{2h}\,\xxi_k,\
\bL^E_0=0,
\end{equation}
for $k=0,1,\ldots,[T/h]-1,$ where $\xxi_1$, $\xxi_2,\ldots$ are
i.i.d.\ standard Gaussian random vectors in $\RR^M$ and $[x]$ stands
for the integer part of $x\in\RR$. Obviously, the sequence
$\{\bL_k^E;k\ge 0\}$ defines a discrete-time Markov process.
Furthermore, one can show that this Markov process can be
extrapolated to a continuous-time diffusion-type process which
converges in distribution to the Langevin diffusion as $h\to 0$.
Here extrapolation means the construction of a process $\{\tilde
\bL_{t,h};t\in[0,T]\}$ satisfying $\tilde \bL_{kh,h} =\bL_k^E$ for
every $k=0,\ldots,[T/h]$. Such a process $\{\tilde
\bL_{t,h};t\in[0,T]\}$ can be defined as a solution of the SDE
$$
d\tilde \bL_{t,h}=\!\!\sum_{k=0}^{[T/h]-1}\!\!
\1_{[k,k+1)}(t/h)\nabla V(\bL_k^E)\,d t+\sqrt{2}\,d\bW_t,\ t\ge 0.
$$
This amounts to connecting the successive values of the Mar\-kov
chain by independent Brownian bridges. The Girsanov formula implies
that the Kullback-Leibler divergence between the distribution of the
process $\{\bL_t;t\in[0,T]\}$ and the distribution of
$\{\tilde\bL_{t,h}\!\!\!\!\!\!\!\!{\phantom{L_t}}^E;t\in[0,T]\}$
tends to zero as $h$ tends to zero. Therefore, it makes sense to
approximate $\bar\bL_T$ by
$$
\bar\bL_{T,h}^E=\frac{h}{T}\sum_{k=0}^{[T/h]-1} \bL_k^E.
$$
\begin{proposition}\label{PropConv}
Consider the linear model $\bY= \mathbb X\llambda^* + \xxi$, where
$\mathbb X$ is the $n\times M$ deterministic matrix and $\xxi$ is a
zero-mean noise with finite covariance matrix. Then for
$\hat\llambda=\int_{\RR^M}\llambda\,p_V(\llambda)\,d\llambda$ with
$p_V(\llambda)\propto e^{V(\llambda)}$ and $V(\llambda)$ defined in
(\ref{V}) we have
$$
\varlimsup_{T\to\infty}\varlimsup_{h\to 0} \Ex\Big[\big\|\bar\bL_{T,h}^E -\hat\llambda\big\|_2\Big]=0.
$$
\end{proposition}

\begin{proof}
We present here a high-level overview of the proof deferring the
details to the Appendix.
\begin{description}
    \item[Step 1] We start by showing that
    $$
    \lim_{h\to 0}\Ex\bigg\|\bar\bL_{T,h}^E-
    \frac1T\int_0^T \tilde \bL_{t,h}\,dt\bigg\|_2^2=0.
    $$
\item[Step 2] We then split the expression $\frac1T\int_0^T \tilde \bL_{t,h}\,dt$   into two terms:
    \begin{equation}\label{eq1}
    \frac1T\int_0^T \tilde \bL_{t,h}\,dt=\underbrace{\frac1T\int_0^T \tilde \bL_{t,h}\1_{[0,A]}(\|\tilde \bL_{t,h}\|_2)\,dt}_{\textsf{T}_1}
    +\underbrace{\frac1T\int_0^T \tilde \bL_{t,h}\1_{]A,+\infty]}(\|\tilde \bL_{t,h}\|_2)\,dt}_{\textsf{T}_2}.
    \end{equation}
    and show that the expected norm $\Ex\|\textsf{T}_2\|_2$ is bounded uniformly in $h$ and $T$ by some function of $A$ that decreases to 0
    as $A\to\infty$. Later $A$ will be chosen as an increasing function of $T$.

    \item[Step 3] We check that the Kullback-Leibler divergence between the distribution of $(\tilde \bL_{t,h};0\le t\le T)$
    and of
    $(\bL_{t};0\le t\le T)$ tends to zero as $h\to 0$. This implies the convergence in total variation and, as a consequence,
    we get
    \begin{equation}\label{eq2}
    \lim_{h\to 0}\Ex\bigg[\bigg(\frac{1}{T}\int_0^T\!\! G(\tilde \bL_{t,h})\,dt-
    \int_{\RR^M}\!\! G(\llambda)p_V(\llambda)d\llambda\bigg)^2\bigg]=
    \Ex\bigg[\bigg(\frac{1}{T}\int_0^T\!\! G(\bL_{t})\,dt-
    \int_{\RR^M}\!\! G(\llambda)p_V(\llambda)d\llambda\bigg)^2\bigg]
    \end{equation}
    for any bounded measurable function $G:\RR^M\to\RR$. We
    use this result with $G(\llambda)=\lambda_i\1_{[0,A]}(\|\llambda\|_2)$,
    $i=1,\ldots,M$.

    \item[Step 4] To conclude the proof we use the fact that
    $\int_{\|\llambda\|_2>A}\llambda p_V(\llambda)\,d\llambda$ tends
    to zero as $A\to\infty$, and that by the ergodic theorem (cf.
    Proposition~1) the right hand side of (\ref{eq2}) tends to 0 as
    $T\to\infty$.
    \end{description}
    \vspace{-10pt}
\end{proof}

This discretization algorithm is easily implementable and, for small
values of $h$, $\bar\bL_{T,h}^E$ is very close to the integral
$\hat\llambda=\int \llambda \, p_V(\llambda)\,d\llambda$ of
interest. However, for some values of $h$, which may eventually be
small but not enough, the Markov process $\{\bL_k^E; k\ge 0\}$ is
transient. Therefore, if $h$ is not small enough the sum in the
definition of $\bar\bL_{T,h}^E$ explodes \cite{RobStr02}. To circumvent this
problem, one can either modify the Markov chain $\bL_k^E$ by
incorporating a Metropolis-Hastings correction, or take
a smaller $h$ and restart the computations. The Metropolis-Hastings approach
guarantees the convergence to the desired distribution. However, it
considerably slows down the algorithm because of a significant
probability of rejection at each step of discretization. The second
approach, where we just take a smaller $h$, also slows down the
algorithm but we keep some control on its time of execution.

\section{Implementation and experimental results}\label{S5}

In this section we give more details on the implementation of the
LMC for computing the EW-aggregate in the linear regression model.

\subsection{Implementation}
The input of the algorithm we are going to describe is the triplet
$(\bY,\mathbb X,\sigma)$ and the tuning parameters
$(\alpha,\beta,\tau,h,T)$, where \vglue-5pt \vbox{
\begin{itemize}
\item[-] $\bY$ is the $n$-vector of values of the response variable, \\[-19pt]
\item[-] $\mathbb X$ is the $n\times M$ matrix of predictor variables,\\[-19pt]
\item[-] $\sigma$ is the noise level,\\[-19pt]
\item[-] $\beta$ is the temperature parameter of the EW-aggregate,\\[-19pt]
\item[-] $\alpha$ and $\tau$ are the parameters of the sparsity prior,\\[-19pt]
\item[-] $h$ and $T$ are the parameters of the LMC algorithm.
\end{itemize}
} \vskip-5pt The output of the proposed algorithm is a vector
$\hat\llambda\in\RR^M$ such that, for every $\bx\in\RR^M$,
$\bx^\T\hat\llambda$ provides a prediction for the unobservable
value of the response variable corresponding to $\bx$. The
pseudo-code of the algorithm is given below. \vglue-5pt

\begin{algorithm}[ht!]
\SetLine\small
\boxed{
\vbox{
\KwIn{\small Observations $(\bY,\mathbb X,\sigma)$ and parameters $(\alpha,\beta,\tau,h,T)$}
\KwOut{The vector $\hat\llambda$}
Set \\
\texttt{\noindent\small
\hskip25pt [n,M]=size(X)\;
\hskip25pt L=zeros(M,1)\;
\hskip25pt lambda=zeros(M,1)\;
\hskip25pt H=0\;
}
Calculate \\
\texttt{\noindent\small
\hskip25pt XX=X'*X\;
\hskip25pt Xy=X'*y\;
}

\While{ {\tt H} is less than {\tt T} }{
 \texttt{\small nablaV=(2/$\beta$)*(Xy-XX*L)-$\alpha$*$\bar\omega'(\alpha$L$)$\;
 nablaV=nablaV-4*L./($\tau$\^{}2+L.\!\^{}2)\;
 L=L+h*nablaV+sqrt(2*h)*randn(M,1)\;
 H=H+h\;
 lambda=lambda+h*L/T\;}
}
\textbf{return} \texttt{lambda}
}}
\smallskip
\caption{The algorithm for computing the EW-aggregate by LMC.}
\label{alg:mine}
\end{algorithm}

\vglue-60pt

\begin{description}
\item[Choice of $T$:]
Since the convergence rate of $\bar\bL_T$ to $\hat\llambda$ is of
the order $T^{-1/2}$ and the best rate of convergence an estimator
can achieve is $n^{-1/2}$, it is natural to set $T=n$. This choice
of $T$ has the advantage of being simple for implementation, but it
has the drawback of  being not scale invariant. A better strategy
for choosing $T$ is to continue the procedure until the convergence
is observed.

\item[Choice of $h$:]
We choose the step of discretization in the form: $h=\beta/({Mn})=\beta/{\tr(\mathbb X^\T \mathbb X)}$.
More details on the choice of $h$ and $T$ will be given in
a future work.

\item[Choice of\ \ $\beta$, $\tau$ and $\alpha$:]
In our simulations we use the parameter values
$$
\alpha=0,\qquad \beta=4\sigma^2,\qquad
\tau={4\sigma}/{({\tr(\mathbb X^\T \mathbb X)})^{1/2}}   \ .
$$
These values of $\beta$ and $\tau$ are derived from the theory
developed above. However, we take here $\alpha=0$ and not $\alpha>0$
as suggested in Section \ref{S3}. We introduced there $\alpha>0$ for
theoretical convenience, in order to guarantee the geometric mixing
of the Langevin diffusion. Numerous simulations show that mixing
properties of the Langevin diffusion are preserved with $\alpha=0$
as well.
\end{description}

\subsection{Numerical experiments}
We present below two examples of application of the EWA with LMC for
simulated data sets. In both examples we give also the results
obtained by the Lasso procedure (rather as a benchmark, than for
comparing the two procedures). The main goal of this section is to
illustrate the predictive ability of the EWA and to show that it can
be easily computed for relatively large dimensions of the problem.
In all examples the Lasso estimators are computed with the
theoretically justified value of the regularaization parameter
$\sigma\sqrt{8\log M/n}$ (cf. \cite{BRT08}).

\subsubsection{Example 1}
This is a standard numerical example where the Lasso and Dantzig
selector are known to behave well (cf. \cite{CT07}). Consider the
model $\bY=\mathbb X\llambda^*+\sigma\xxi$, where $\mathbb X$ is a
$M\times n$ matrix with independent entries, such that each entry is
a Rademacher random variable. Such matrices are particularly well
suited for applications in compressed sensing. The noise
$\xxi\in\RR^n$ is a vector of independent standard Gaussian random
variables. The vector $\llambda^*$ is chosen to be $S$-sparse, where
$S$ is much smaller than $M$. W.\,l.\,o.\,g.\ we consider vectors
$\llambda^*$ such that only first $S$ coordinates are different from
$0$; more precisely, $\lambda_j^*=\1(j\le S)$. Following
\cite{CT07}, we choose $\sigma^2={S}/9$. We run our procedure for
several values of $S$ and $M$. The results of 500 replications are
summarized in Table~1. We see that EWA outperforms Lasso in all the
considered cases.

\begin{figure}[t] %
\centerline{\includegraphics[width=430pt]{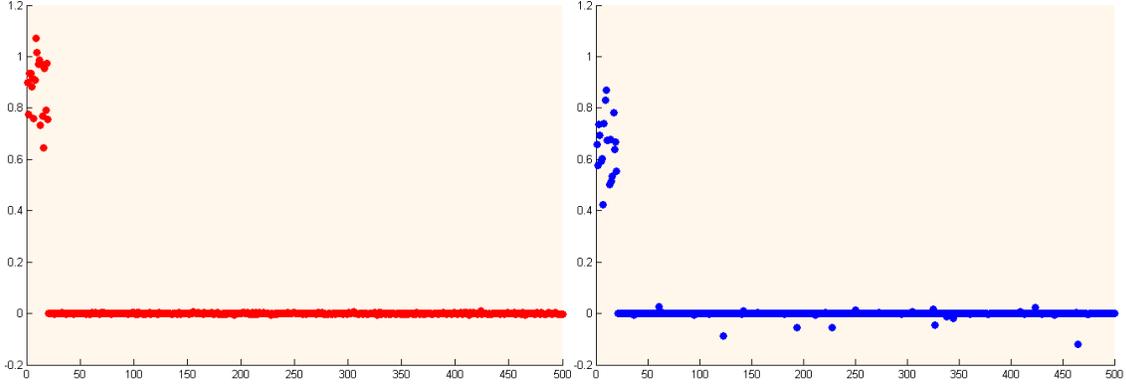}}
\vglue-5pt
\caption[]{A typical result of the EWA (left panel) and the Lasso
(right panel) in the setup of Example 1 with $n=200$, $M=500$ and
$S=20$.} \label{fig-2}
\end{figure}

A typical scatterplot of
estimated coefficients for $M=500$, $n=200$ and $S=20$ is presented
in Fig.~\ref{fig-2}. The left panel shows the estimated coefficients
obtained by EWA, while the right panel shows the estimated coefficients
obtained by Lasso. One can clearly see that the estimated values provided
by EWA are much more accurate than those provided by Lasso.

An interesting observation is that the EWA selects the set of
nonzero coordinates of $\llambda^*$ even better than the Lasso does.
In fact, the approximate sparsity of the EWA is not very surprising,
since in the noise-free linear models with orthogonal matrix
$\mathbb X$, the symmetry of the prior implies that the EWA recovers
the zero coordinates without error.

We note that the numerical results on the Lasso in
Table~\ref{table1} are substantially different from those reported
in the short version of this paper published in the Proceeding of
COLT 2009 \cite{DT09}. This is because in \cite{DT09} we used the R
packages \texttt{lars} and \texttt{glmnet}, whereas here we use the
MATLAB package \PVerb{l1_ls}. It turns out that in the present example
the latter provides more accurate approximation of the Lasso than
the aforementioned R packages.

The running times of our algorithm are reasonable. For instance, in
the case $n=m=100$ and $S=10$ the execution of our algorithm is only
three times longer than the \PVerb{l1-ls} implementation of the
Lasso. On the other hand, the prediction error of our algorithm is
more than twice smaller than that of the Lasso.

\begin{table}\label{table1}
\begin{center}

\begin{tabular}{c|cc|cc|cc}
\toprule
& &{$M=100$} & &{$M=200$} & &{$M=500$}\\
 & \textbf{EWA} & \textbf{Lasso} & \textbf{EWA} & \textbf{Lasso} & \textbf{EWA} & \textbf{Lasso}\\
\toprule
$n=100$ $S=5$  & 0.063 & 0.344 & 0.064 & 0.385 & 0.087 &  0.453 \\
   & (0.039)& (0.132) & (0.043) & (0.151) & (0.054) &  (0.161) \\
%   & $T=2$ &  & $T=1$ &  & $T=1$ & \\
\hline
$n=100$ $S=10$ & 0.73725 & 1.680 & 1.153 & 1.918 & 1.891 & 2.413 \\
          & (0.699)& (0.621)& (1.091)& (0.677)& (1.522)& (0.843)\\
%         & $T=1$ &  & $T=2$ &  & $T=5$ & \\
\hline
$n=100$ $S=15$ &5.021 & 4.330 & 6.495   & 5.366 & 8.917 &  7.1828 \\
          & (1.593) & (1.262) & (1.794) & (1.643)& (2.186) &  (2.069)\\
%   & $T=10$ &  & $T=20$ &  & $T=10$ & \\
\toprule
$n=200$ $S=5$ & 0.021 & 0.151 & 0.022 &  0.171 & 0.019 & 0.202 \\
          & (0.011)&  (0.048)& (0.013)& (0.055) & (0.012)& (0.057)\\
%   & $T=1$ &  & $T=1$ &  & $T=1$ & \\
\hline
$n=200$ $S=10$ & 0.106 & 0.658 & 0.108 & 0.753 & 0.117 & 0.887 \\
        & (0.047)& (0.169) & (0.048) & (0.198) & (0.051) & (0.239) \\
%   & $T=2$ &  & $T=10$ &  & $T=2$ & \\
\hline
$n=200$ $S=20$ & 1.119 & 3.124 & 1.6015 & 3.734 & 2.728  & 4.502 \\
         & (0.696) & (0.806)& (1.098) & (0.907) & (1.791)& (1.063)\\
%   & $T=10$ &  & $T=10$ &  & $T=20$ & \\
\hline
\end{tabular}\end{center}
\vglue-14pt
\caption
{Average loss $\|\hat\llambda-\llambda^*\|^2$ of the
estimators obtained by the EW-aggregate and the Lasso in Example 1. The
standard deviation is given in parentheses.}
\end{table}

\subsubsection{Example 2}
Consider model (\ref{model}) where $Z_i$ are independent random
variables uniformly distributed in the unit square $[0,1]^2$ and
$\xi_i$ are iid ${\mathcal N}(0,\sigma^2)$ random variables. For an
integer $k>0$, we consider the indicator functions of rectangles
with sides parallel to the axes and having as left-bottom vertex the
origin and as right-top vertex a point of the form $(i/k,j/k)$,
$(i,j)\in\mathbb N^2$. Formally, we define $\phi_j$ by
$$
\phi_{(i-1)k+j}(x)=\1_{[0,i]\times[0,j]}(kx),\qquad\forall\,x\in[0,1]^2.
$$
The underlying image $f$ we are trying to recover is taken as a
superposition of a small number of rectangles of this form, that is
$f(x)=\sum_{\ell=1}^{k^2} \lambda_\ell^*\phi_\ell(x)$, for all
$x\in[0,1]^2$ with some $\llambdastar$ having a small
$\ell_0$-norm. We set $k=15$, $\|\llambdastar\|_0=3$,
$\lambda_{10}^*=\lambda_{100}^*=\lambda_{200}^*=1$. Thus, the
cardinality of the dictionary is $M=k^2=225$.
%Before computing the EW-aggregate and
%the Lasso estimate, all the covariates have been normalized.

In this example the functions $\phi_j$ are strongly correlated and
therefore the assumptions like restricted isometry or low coherence
are not fulfilled. Nevertheless, the Lasso succeeds in providing an
accurate prediction (cf.\ Table~2). Furthermore, the Lasso with the
theoretically justified choice of the regularization parameter
$\sigma\sqrt{8\log M/n}$ is not much worse than the ideal
Lasso-Gauss (LG) estimator. We call the LG estimator the ordinary
least squares estimator in the reduced model where only the
predictor variables selected at a preliminary Lasso step are kept.
Of course, the performance of the LG procedure depends on the
initial choice of the tuning parameter for the Lasso step. In our
simulations, we use its ideal (oracle) value minimizing the
prediction error and, therefore, we call the resulting procedure the
ideal LG estimator.
\begin{table}[t]
\label{table2}
\begin{center}
\begin{tabular}{c|ccc}
\toprule
 & \textbf{EWA} & \textbf{Lasso}  & \textbf{Ideal LG} \\
\toprule
$\sigma=1,$ $n=100$ & 0.160 & 0.273  & 0.128   \\
%$T=5$
& (0.035)& (0.195) & (0.053)\\
\hline
$\sigma=2,$ $n=100$ & 0.210 & 0.759  & 0.330   \\
%$T=1$
& (0.072)& (0.562) & (0.145)\\
\hline
$\sigma=4,$ $n=100$ & 0.420 & 2.323 & 0.938 \\
%$T=1$
& (0.222)& (1.257) & (0.631) \\
\hline
$\sigma=1,$ $n=200$ & 0.130  & 0.187  & 0.069   \\
%$T=6$
& (0.030)& (0.124) & (0.031)\\
\hline
$\sigma=2,$ $n=200$ & 0.187 & 0.661 & 0.203  \\
%$T=1$
& (0.048) & (0.503) & (0.086) \\
\hline
$\sigma=4,$ $n=200$ & 0.278 & 2.230 & 0.571 \\
%$T=1$
& (0.132) & (1.137) & (0.324)  \\
\hline
\end{tabular}
\end{center}
\vglue-5pt
\caption{Average loss $\int_{[0,1]^2}\big(\sum_{j}(\hat\lambda_j-\lambda_j^*)\phi_j(x)\big)^2\,dx$ of the
the EWA, the Lasso and the ideal LG procedures in Example 2. The
standard deviation is given in parentheses.}
\end{table}

As expected, the EWA has a smaller predictive risk than the Lasso
estimator. However, a surprising outcome of this experiment is the
supremacy of the EWA over the ideal LG in the case of large noise
variance. Of course, the LG procedure is
 faster. However, even from this point of view the EWA is rather attractive, since it
takes less than two seconds to compute it in the present example.

\begin{figure}[t] %

\centerline{\includegraphics[width=0.85\linewidth]{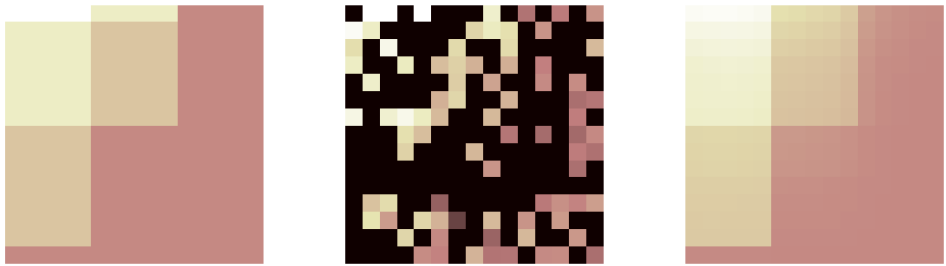}}
\vglue-5pt
\caption[]{This figure shows a typical outcome in the setup of example 2 when $n=200$ and $k=15$.
\textit{Left:} the original image. \textit{Center:} the observed noisy sample with $\sigma=0.5$. Pixels
for which no observation is available are in black. \textit{Right:} the image estimated by the EWA.}
\label{fig3}
\end{figure}

\section{Conclusion and outlook}\label{S6}

This paper contains two contributions: New oracle inequalities for
EWA, and the LMC method for approximate computation of the EWA. The
first oracle inequality presented in this work is in the line of the
PAC-Bayesian bounds initiated by McAllester \cite{McAllester03}. It
is valid for any prior distribution and gives a bound on the risk of
the EWA with an arbitrary family of functions. Next, we derive
another inequality, which is adapted to the sparsity scenario and
called the sparsity oracle inequality (SOI). In order to obtain it,
we propose a prior distribution favoring sparse representations. The
resulting EWA is shown to behave almost as well as the best possible
linear combination within a residual term proportional to $M^*(\log
M)/n$, where $M$ is the true dimension, $M^*$ is the number of atoms
entering in the best linear combination and $n$ is the sample size.
A remarkable fact is that this inequality is obtained under no
condition on the relationship between different atoms.

Sparsity oracle inequalities similar to that of
Theorem~\ref{ThmSparsity} are valid for the penalized empirical risk
minimizers (ERM) with a $\ell_0$-penalty (proportional to the number
of atoms involved in the representation). It is also well known that
the problem of computing the $\ell_0$-penalized ERM is NP-hard. In
contrast with this, we have shown that the numerical evaluation of
the suggested EWA is a computationally tractable problem. We
demonstrated that it can be efficiently solved by the LMC algorithm.
Numerous simulations we did (some of which are included in this
work) confirm our theoretical findings and, furthermore, suggest
that the EWA is able to efficiently select the sparsity pattern.
Theoretical justification of this fact, as well as more thorough
investigation of the choice of parameters involved in the LMC
algorithm, are interesting topics for future research.

\section*{Appendix: proofs of technical results}

\subsection{Proof of Proposition~\ref{PropConv}}

%To ease notation and since no confusion is possible, throughout this proof the Euler discretization
%of the Langevin diffusion will be denoted by $(\bL_{k};k=0,1,2,\ldots)$ instead of $(\bL_{k}^E;k=0,1,2,\ldots)$.
For brevity, in this proof we denote by $\|\cdot\|$ the Euclidean
norm in $\RR^M$ and we set $\alpha=1$ in (\ref{V}). The case of
general $\alpha>0$ is treated analogously. Recall that for some
small $h>0$ we have defined the $M$-dimensional Markov chain
$(\bL^E_{k};k=0,1,2,\ldots)$ by (cf. (\ref{V}) and (\ref{CSE})):
$$
\bL^E_{k+1}=\bL^E_k+2h\beta^{-1}\bbX^\T(\bY-\bbX
\bL^E_{k})-hg(\bL^E_k)+\sqrt{2h}\,\xxi_{k+1},\quad \bL^E_0=0,
$$
where $(\xxi_k;k=1,2,\ldots)$ is a sequence of iid standard Gaussian
vectors in $\RR^M$, and
$$
g:\RR^M\to\RR^M\quad\text{s.t.}\quad
g(\llambda)=\bigg(\frac{4\lambda_1}{\tau^2+\lambda_1^2}+\bar\omega'(\lambda_1),\ldots,\frac{4\lambda_M}{\tau^2+\lambda_M^2}+\bar\omega'(\lambda_M)\bigg)^\T.
$$
In what follows, we will use the fact that the function $g$ is bounded and satisfies $\llambda^\T g(\llambda)\ge 0$ for all $\llambda\in\RR^M$.

Let us prove some auxiliary results. Set $\bv=2\beta^{-1}\bbX^\T
\bY$, $\ttA=2\beta^{-1}\bbX^\T\bbX$ and assume that $h\le
1/\|\ttA\|$. Without loss of generality we also assume that $T/h$ is
an integer. In what follows, we denote by $C>0$ a constant whose
value is not essential, does not depend neither on $T$ nor on $h$,
and may vary from line to line.  Since the function $g$ is bounded
and $\xxi_{k+1}$ has zero mean, we have
$$
\Ex[\bL^E_{k+1}]=(I-h\ttA)\Ex[\bL^E_k]+h\Ex[\bv-g(\bL^E_k)],\qquad
\forall k\ge 0.
$$
Therefore,
\begin{equation*}
\|\Ex[\bL^E_{k+1}]\|\le \|(I-h\ttA)\Ex[\bL^E_k]\|+Ch\le
\|\Ex[\bL^E_k]\|+Ch,\qquad \forall k\ge 0.
\end{equation*}
By induction, we get
\begin{equation}\label{eq3}
\|\Ex[\bL^E_{k}]\|\le Ckh\le CT,\qquad \forall k\in[0,[T/h]].
\end{equation}
Furthermore, since $\xxi_{k+1}$ is independent of $\bL^E_k$ and $Y$,
we have
\begin{eqnarray*}
\Ex[\|\bL^E_{k+1}\|^2]&=&\Ex[\|\bL^E_k+h\bv -h\ttA \bL^E_k-hg(\bL^E_k)\|^2]+2hM \nonumber\\
&\le& \Ex[\|\bL^E_k\|^2+2h (\bL^E_k)^\T (\bv -\ttA \bL^E_k)-2h(\bL^E_k)^\T g(\bL^E_k)+h^2\|\bv -\ttA \bL^E_k-g(\bL^E_k)\|^2]+2hM\nonumber\\
&\le& \Ex[\|\bL^E_k\|^2+2h (\bL^E_k)^\T (\bv -\ttA \bL^E_k)+2h^2\|\ttA \bL^E_k\|^2+2h^2\|\bv -g(\bL^E_k)\|^2]+2hM\nonumber\\
&\le& \Ex[\|\bL^E_k\|^2+2h (\bL^E_k)^\T \bv -2h(\bL^E_k)^\T(\ttA-h\ttA^2)\bL^E_k+2h^2\|\bv -g(\bL^E_k)\|^2]+2hM\nonumber\\
&\le& \Ex[\|\bL^E_k\|^2]+2h \Ex[(\bL^E_k)^\T] \bv+Ch\nonumber\\
&\le& \Ex[\|\bL^E_k\|^2]+ChT,\qquad \forall k\in[0,[T/h]].
\end{eqnarray*}
Once again, using induction, we get
\begin{equation}\label{eq4}
\Ex[\|\bL^E_k\|^2]\le CkhT\le CT^2,\qquad \forall k\in[0,[T/h]].
\end{equation}
This implies, in particular, that $(h/T)\Ex[\|\bL^E_{[T/h]}\|^2]\to
0$ as $h\to 0$ for any fixed $T$.

\bigskip
\paragraph{\textbf{Proof of Step 1}} Denote by $\psi$ the function
$$
\psi(\llambda)=\bv-\ttA \llambda-g(\llambda),\qquad
\forall \llambda\in\RR^M,
$$
and define the continuous-time random process $(\tilde \bL_{t,h};
0\le t\le [T/h]h)$ by
\begin{equation}\label{eq5}
d\tilde
\bL_{t,h}=\sum_{k=0}^{[T/h]-1}\psi(\bL^E_k)\1_{[kh,(k+1)h)}(t)\,dt+\sqrt{2}d\bW_t,\quad
\tilde \bL_{0,h}=\boldsymbol 0,
\end{equation}
where $\bW_t$ is a $M$-dimensional Brownian motion satisfying $\bW_{kh}=\xxi_k$, for all $k$. The rigorous construction of
$\bW$ can be done as follows. Let $(\bB_t;0\le t\le T)$ be a $M$-dimensional Brownian motion defined on the same probability space
as the sequence $(\xxi_k;0\le k\le [T/h])$ and independent of $(\xxi_k;0\le k\le [T/h])$. One can check that the process defined by
$$
\bW_t=\xxi_k+\bB_t-\bB_{kh}-\Big(\frac{t}{h}-k\Big)\big(\bB_{(k+1)h}-\bB_{kh}-\xxi_{k+1}\big),\qquad t\in [kh,(k+1)h[
$$
is a Brownian motion and satisfies $\bW_{kh}=\xxi_k$.

By the Cauchy-Schwarz inequality,
\begin{eqnarray*}
    \Ex\bigg[\bigg\|\frac{h}{T}\sum_{k=0}^{[T/h]-1} \bL^E_k-
    \frac1T\int_0^T \tilde \bL_{t,h}\,dt\bigg\|^2\bigg]&=&
    \Ex\bigg[\bigg\|\frac1T\int_0^T
    \big(\sum_{k=0}^{[T/h]-1}\tilde \bL_{kh,h}\1_{[kh,(k+1)h)}(t)-\tilde \bL_{t,h}\big)\,dt\bigg\|^2\bigg]\\
    &\le& \frac1T \sum_{k=0}^{[T/h]-1} \int_{kh}^{(k+1)h} \Ex\big[\|\tilde \bL_{t,h}-\tilde \bL_{kh,h}\|^2\big]\,dt\\
    &\le& \frac2T \sum_{k=0}^{[T/h]-1} \int_{kh}^{(k+1)h}\limits \Ex\big[h^2\|\psi(\bL^E_{k})\|^2+2\|\bW_t-\bW_{kh}\|^2\big]\,dt\\
    &\le& \frac{2h^3}T \sum_{k=0}^{[T/h]-1} \Ex\big[\|\psi(\bL^E_{k})\|^2\big]+4Mh.
\end{eqnarray*}
Using the inequality $\|\psi(\llambda)\|\le C(1+\|\llambda\|)$ and
(\ref{eq4}), we get
\begin{eqnarray*}
    \Ex\bigg[\bigg\|\frac{h}{T}\sum_{k=0}^{[T/h]-1} \bL^E_k- \frac1T\int_0^T \tilde \bL_{t,h}\,dt\bigg\|^2\bigg]
    &\le& Ch^2 +\frac{Ch^3}{T}\sum_{k=0}^{[T/h]-1} \Ex\big[\|\bL^E_{k}\|^2\big]+4Mh\\
    &\le& Ch(1+hT^2).
\end{eqnarray*}
This completes the proof of the Step 1.

\bigskip
\paragraph{\textbf{Proof of Step 2}}
Using (\ref{eq4}) we obtain
\begin{eqnarray}
\Ex[\|\textsf{T}_2\|]&\le& \frac1T\int_0^T\Ex[\|\tilde \bL_{t,h}\|\1_{[A,+\infty]}(\|\tilde \bL_{t,h}\|)]\le\frac{1}{TA}\int_0^T \Ex[\|\tilde \bL_{t,h}\|^2]\,dt\nonumber\\
&\le&\frac{C}{TA}\sum_{k=0}^{[T/h]-1}\int_{kh}^{(k+1)h} \big(\Ex[\|\bL^E_k\|^2]+
h^2\Ex[\|\psi(\bL^E_k)+\ttA \bL^E_k\|^2]+\Ex[\|\bW_{t}-\bW_{kh}\|^2]\big)\,dt\nonumber\\
&\le&\frac{C}{TA}\sum_{k=0}^{[T/h-1]} h\big(\Ex[\|\bL^E_k\|^2]+C
h^2+Mh\big)\le \frac{CT^2}{A}.
\end{eqnarray}
Thus, choosing, for example, $A=T^3$ we guarantee that
$\lim_{T\to\infty}\varlimsup_{h\to 0}\Ex[\|\textsf{T}_2\|]=0$.

\bigskip
\paragraph{\textbf{Proof of Step 3}}
First, note that (\ref{eq5}) can be written in the form
\begin{equation*}
d\tilde \bL_{t,h}=\tilde\psi(\tilde \bL_{h},t)\,dt+\sqrt{2}d\bW_t,\quad \tilde \bL_{0,h}=\boldsymbol 0,
\end{equation*}
where $\tilde\psi(\tilde\bL_h,t)$ is a non-anticipative process that
equals $\psi(\tilde\bL_{kh,h})$ when $t\in[kh,(k+1)h)$. Recall that
the Langevin diffusion is defined by the stochastic differential
equation
\begin{equation*}
d\bL_{t}=\psi(\bL_t)\,dt+\sqrt{2}d\bW_t,\quad \bL_{0}=\boldsymbol 0.
\end{equation*}
Therefore, the probability distributions $\Pb_{\bL,T}$ and
$\Pb_{\tilde\bL_h,T}$ induced by, respectively, $(\bL_t;0\le t\le
T)$ and $(\tilde\bL_{t,h};0\le t\le T)$ are mutually absolutely
continuous and the corresponding Radon-Nykodim derivatives are given
by Girsanov formula:
$$
\frac{d\Pb_{\tilde\bL_h,T}}{d\Pb_{\bL,T}}(\bL)=\exp\Big\{\frac1{\sqrt{2}}\int_0^T (\tilde\psi(\bL,t)-\psi(\bL_t))^\T \big(d\bL_t-
\psi(\bL_t)\,dt\big)-\frac14\int_0^T \|\tilde\psi(\bL,t)-\psi(\bL_t)\|^2\,dt\Big\}.
$$
This implies that the Kullback-Leibler divergence between $\Pb_{\tilde\bL_{h},T}$ and $\Pb_{\bL,T}$ is given by
$$
\KL\big(\Pb_{\bL,T}|\!|\Pb_{\tilde\bL_{h},T}\big)=-\Ex\Big[\log\Big(\frac{d\Pb_{\tilde\bL_h,T}}{d\Pb_{\bL,T}}(\bL)\Big)\Big]=\frac14\int_0^T\Ex\big[\|\tilde\psi(\bL,t)-\psi(\bL_t)\|^2\big]\,dt.
$$
Using the expressions of $\psi$ and $\tilde\psi$, as well as the fact that the function $\psi$ is Lipschitz continuous, we can bound the divergence above as follows:
\begin{eqnarray*}
\KL\big(\Pb_{\bL,T}|\!|\Pb_{\tilde\bL_{h},T}\big)&=&\frac14\sum_{k=0}^{[T/h]-1}\int_{kh}^{(k+1)h}\Ex\big[\|\psi(\bL_{kh})-\psi(\bL_t)\|^2\big]\,dt\\
&\le& C \sum_{k=0}^{[T/h]-1}\int_{kh}^{(k+1)h}\Ex\big[\|\bL_{kh}-\bL_t\|^2\big]\,dt\\
&=& C \sum_{k=0}^{[T/h]-1}\int_{kh}^{(k+1)h}\Ex\Big[\Big\|\int_{kh}^t \psi(\bL_s)\,ds+\sqrt{2}(\bW_t-\bW_{kh})\Big\|^2\Big]\,dt.
\end{eqnarray*}
From the Cauchy-Schwarz inequality and the fact that
$\|\psi(\llambda)\|\le C(1+\|\llambda\|)$ we obtain
\begin{eqnarray*}
\KL\big(\Pb_{\bL,T}|\!|\Pb_{\tilde\bL_{h},T}\big)
&\le& C \sum_{k=0}^{[T/h]-1}\int_{kh}^{(k+1)h} h\int_{kh}^t
\Ex\big[\|\psi(\bL_s)\|^2\big]\,ds\,dt+ChT\\
&\le& Ch^2 \sum_{k=0}^{[T/h]-1}\int_{kh}^{(k+1)h}
\Ex\big[\|\psi(\bL_s)\|^2\big]\,ds+ChT\\
&\le& C h^2 \int_{0}^{T} \Ex\big[\|\psi(\bL_s)\|^2\big]\,ds+ChT\\
&\le& C h^2 \int_{0}^{T} \Ex\big[\|\bL_s\|^2\big]\,ds+ChT.
\end{eqnarray*}
Since by Proposition~\ref{PropDrift} the expectation of
$\|\bL_s\|^2$ is bounded uniformly in $s$, we get
$\KL\big(\Pb_{\bL,T}|\!|\Pb_{\tilde\bL_{h},T}\big)\to 0$ as $h\to
0$. In view of Pinsker's inequality, cf, e.g., \cite{Tsyb09}, this
implies that the distribution $\Pb_{\tilde\bL_{h},T}$ converges to
$\Pb_{\bL,T}$ in total variation as $h\to 0$. Thus, (\ref{eq2})
follows.

\bigskip
\paragraph{\textbf{Proof of Step 4}} To prove that the right hand
side of (\ref{eq2}) tends to zero as $T\to +\infty$, we use the fact
that the process $\bL_t$ has the geometrical mixing property with
$D(\llambda)=e^{\alpha\|\llambda\|_2}$. Bias-variance decomposition
yields:
\begin{eqnarray*}
    \Ex\bigg[\bigg(\frac{1}{T}\int_0^T\!\! G(\bL_{t})\,dt-
    \int G(\llambda) p_V(\llambda)d\llambda\bigg)^2\bigg]=\frac1{T^2}\textbf{Var}\Big[\int_0^T\!\! G(\bL_t)\,dt\Big]+
    \bigg(\frac{1}{T}\int_0^T \Ex_0[G(\bL_{t})]\,dt-\int G(\llambda) p_V(\llambda)d\llambda\bigg)^2\!\!.
\end{eqnarray*}
The second term on the right hand side of the last display tends to
zero as $T\to\infty$ in view of Proposition~\ref{PropDrift}, while
the first term can be evaluated as follows:
\begin{eqnarray*}
    \frac1{T^2}\textbf{Var}\Big[\int_0^T G(\bL_t)\,dt\Big]&=&\frac1{T^2}\int_0^T \int_0^T \textbf{Cov}_0\big[G(\bL_t),G(\bL_s)\big]\,dt\,ds\\
    &\le& \frac{C}{T^2}\int_0^T \int_0^T \rho_V^{-|t-s|}\,dt\,ds\le C T^{-1}.
\end{eqnarray*}
This completes the proof of Proposition~\ref{PropConv}.

%%%%%%%%%%%%%%%%%%%%%%%%%%%%%%%%%%%%%%%%%%%%%%%
%%%%%%%%%%%%%%%%%%%%%%%%%%%%%%%%%%%%%%%%%%%%%%%
%%%%%%%%%%%%%%%%%%%%%%%%%%%%%%%%%%%%%%%%%%%%%%%

\subsection{Proof of Lemma~\ref{lem:0}}

We first prove a simple auxiliary result, cf. Lemma~\ref{lemm:1}
below. Then, the two claims of Lemma~\ref{lem:0} are proved in
Lemmas~\ref{lem:1} and~\ref{lem:2}, respectively.

\begin{lemma}\label{lemm:1}
For every $M\in\mathbb N$ and every $s>M$, the following inequality holds:
$$
\frac1{(\pi/2)^M}\int_{\big\{u:\|u\|_1>s\big\}} \prod_{j=1}^M\frac{du_j}{(1+u_j^2)^2}\le \frac{M}{(s-M)^2}.
$$
\end{lemma}
\begin{proof}
Let $U_1,\ldots,U_M$ be iid random variables drawn from the scaled
Student $t(3)$ distribution having as density the function $u\mapsto
2/\big[\pi(1+u^2)^2\big]$. One easily checks that $\Ex[U_1^2]=1$.
Furthermore, with this notation, we have
$$
\frac1{(\pi/2)^M}\int_{\big\{u:\|u\|_1>s\big\}} \prod_{j=1}^M\frac{du_j}{(1+u_j^2)^2}=\Pb\Big(\sum_{j=1}^M |U_j|\ge s\Big).
$$
In view of Chebyshev's inequality the last probability can be
bounded as follows:
$$
\Pb\Big(\sum_{j=1}^M |U_j|\ge
s\Big)\le\frac{M\Ex[U_1^2]}{(s-M\Ex[|U_1|])^2}\le
\frac{M}{(s-M)^2}
$$
and the desired inequality follows.
\end{proof}

\begin{lemma}\label{lem:1}
Let the assumptions of Theorem~\ref{ThmSparsity} be satisfied and
let $p_0$ be the probability measure defined by (\ref{eq:6}). If
$M\ge 2$ then
$$
\int_\Lambda (\lambda_1-\lambda_1^*)^2p_0(d\llambda)\le 4\tau^2 e^{4M\alpha\tau}.
$$
\end{lemma}
\begin{proof}
Using the change of variables $u=(\llambda-\llambdastar)/\tau$ we
write
$$
\int_\Lambda (\lambda_1-\lambda_1^*)^2p_0(d\llambda)=
C_{M}\tau^2\int_{B_1(2M)} u_1^2\Big(\prod_{j=1}^M (1+u_j^2)^{-2}e^{-\opi(\alpha\tau u_j)}\Big)
\,du
$$
with
\begin{eqnarray}\label{C_M}
C_M=\Big(\int_{B_1(2M)}\Big(\prod_{j=1}^M
(1+u_j^2)^{-2}e^{-\opi(\tau\alpha u_j)}\Big) \,du\Big)^{-1}
\end{eqnarray}
where $u_j$ are the components of $u$. Bounding the functions
$e^{-\opi(\tau\alpha u_j)}$ by one, extending the integration from
$B_1(2M)$ to $\RR^M$ and using the inequality $\int_\RR
u_1^2(1+u_1^2)^{-2}du_1\le \pi$, we get
$$
\int_\Lambda (\lambda_1-\lambda_1^*)^2p_0(d\llambda)\le
C_{M}\tau^2\pi \Big(\int_{\RR} (1+t^2)^{-2}\,dt\Big)^{M-1}=
2C_{M}\tau^2(\pi/2)^M,
$$
where we used that the primitive of the function $(1+x^2)^{-2}$ is
$\frac12\arctan(x)+\frac{x}{2(1+x^2)}$. To bound $C_M$ we first use
the inequality $\opi(x)\le 2|x|$ which yields:
\begin{equation}\label{inneq}%{eqnarray*}
C_M \le \Big(\int_{B_1(2M)}e^{-2\alpha\tau\|u\|_1}\,\prod_{j=1}^M \frac{du_j}{(1+u_j^2)^2}\Big)^{-1}\\
\le e^{4\alpha\tau M}\Big(\int_{B_1(2M)}\prod_{j=1}^M
\frac{du_j}{(1+u_j^2)^2}\Big)^{-1}.
\end{equation}%{eqnarray*}
In view of (\ref{inneq}) and Lemma~\ref{lem:0} we have %$C_M$ by
%$e^{4\alpha\tau M}(2/\pi)^M\big(1-M/M^2\big)^{-1}$, which leads to
%the estimate
\begin{equation}\label{inneq1}
C_M\le e^{4\alpha\tau M}(2/\pi)^M\big(1-1/M\big)^{-1} \le
2e^{4\alpha\tau M}(2/\pi)^M
\end{equation}%
for $M\ge 2$. Combining these estimates we get
\begin{eqnarray*}
\int_\Lambda (\lambda_1-\lambda_1^*)^2p_0(d\llambda)\le 4\tau^2 e^{4\alpha\tau M}
\end{eqnarray*}
and the desired inequality follows.
\end{proof}

\begin{lemma}\label{lem:2}
Let the assumptions of Theorem~\ref{ThmSparsity} be satisfied and
let $p_0$ be the probability measure defined by (\ref{eq:6}). Then
$$
\KL(p_0,\pi)\le 2\Big(\alpha\|\llambdastar\|_1+\sum_{j=1}^M 2\log(1+|\lj^*|/\tau)\Big)+
(1+4M\alpha\tau).
$$
\end{lemma}
\begin{proof}
The definition of $\pi$, $p_0$ and of the Kullback-Leibler
divergence imply that
\begin{eqnarray}
\KL(p_0,\pi)&=\int_{B_1(2M\tau)} \log\Bigg\{C_M C_{\alpha,\tau,R}\prod_{j=1}^M
\frac{(\tau^2+\lj^2)^2e^{\opi(\alpha\lj)}}{(\tau^2+(\lj-\lj^*)^2)^2
e^{\opi(\alpha(\lj-\lj^*))}}\Bigg\}p_0(d\llambda)\nonumber\\
&=\log(C_M C_{\alpha,\tau,R})+2\sum_{j=1}^M\int_{B_1(2M\tau)} \log\Bigg\{
\frac{\tau^2+\lj^2}{\tau^2+(\lj-\lj^*)^2}\Bigg\} p_0(d\llambda)\nonumber\\
&\qquad
+\sum_{j=1}^M\int_{B_1(2M\tau)}\big(\opi(\alpha\lj)-\opi(\alpha(\lj-
\lj^*))\big)p_0(d\llambda).\label{eq:7}
\end{eqnarray}
We now successively evaluate the three terms on the RHS of
(\ref{eq:7}). First, in view of (\ref{prior}), we have
\begin{eqnarray*}
C_{\alpha,\tau,R}=\int_{B_1(R)} \prod_{j=1}^M \frac{e^{-\opi(\alpha u_j\tau)}}{(1+u_j^2)^2}\,du_j
\le \Big(\int_\RR (1+u_j^2)^{-2}\,du_j\Big)^M=(\pi/2)^M.
\end{eqnarray*}
This and (\ref{inneq1}) imply $\log(C_M C_{\alpha,\tau,R})\le 1+4M\alpha\tau$.

To evaluate the second term on the RHS of (\ref{eq:7}) we use that
\begin{eqnarray*}
\frac{\tau^2+\lambda^2_j}{\tau^2+(\lj-\lj^*)^2}&=& 1+\frac{2\tau(\lj-\lj^*)}{\tau^2+(\lj-\lj^*)^2}(\lj^*/\tau)+\frac{{\lj^*}^2}{\tau^2+(\lj-\lj^*)^2}\\
&\le& 1+|\lj^*/\tau|+(\lj^*/\tau)^2\le (1+|\lj^*/\tau|)^2.
\end{eqnarray*}
This entails that the second term on the RHS of (\ref{eq:7}) is
bounded from above by $\sum_{j=1}^M 2\log(1+|\lj^*|/\tau)$. Finally,
since the derivative of $\opi(\cdot)$ is bounded in absolute value
by $2$, we have $\opi(\alpha\lj)-\opi(\alpha(\lj-\lj^*))\le
2\alpha|\lj^*|$ which implies:
\begin{eqnarray*}
\sum_{j=1}^M\int_{B_1(2M\tau)}\big(\opi(\alpha\lj)-\opi\big(\alpha(\lj-
\lj^*)\big)\big)p_0(d\llambda)\le 2\alpha\|\llambdastar\|_1.
\end{eqnarray*}
Combining these inequalities we get the lemma.
\end{proof}

\subsection{Proofs of remarks~\ref{rem:1}-\ref{rem:6}}
We only prove Remarks \ref{rem:2} and \ref{rem:6}, since the proofs
of the remaining remarks are straghtforward.

\subsubsection{Proof of Remark~\ref{rem:2}}
Let $\xi$ be a random variable satisfying $\Pb(\xi=\pm \sigma)=1/2$ and let $U$ be
another random variable, independent of $\xi$ and drawn from the uniform distribution
on $[-1,1]$. Recall that $\zeta=(1+\gamma)\sigma\sign[\sigma^{-1}\xi-(1+\gamma)U]-\xi$.

We start by proving that $\xi+\zeta$ has the same distribution as
$(1+\gamma)\xi$. Clearly,
$|\xi+\zeta|$ equals $(1+\gamma)\sigma$ almost surely. Furthermore,
\begin{eqnarray*}
\Pb\big(\xi+\zeta=(1+\gamma)\sigma\big)\eqserre\Pb\big(\sigma^{-1}\xi\ge (1+\gamma)U\big)\\
\eqserre\frac12\Big(\Pb\big(1\ge (1+\gamma)U\big)+\Pb\big(-1\ge (1+\gamma)U\big)\Big)\\
\eqserre\frac14\Bigg(\Big(\frac1{1+\gamma}+1\Big)+\Big(-\frac1{1+\gamma}+1\Big)\Bigg)=\frac12.
\end{eqnarray*}
This entails that $\Pb\big(\xi+\zeta=-(1+\gamma)\sigma\big)=1/2$ and, therefore, the
distributions of $\xi+\zeta$ and $(1+\gamma)\xi$ coincide.

We compute now the conditional expectation $\Ex[\zeta|\xi]$. Since $U$ and $\xi$ are
independent, we have
\begin{eqnarray*}
\Ex[\zeta\,|\,\xi=\sigma]\eqserre (1+\gamma)\sigma\Ex\big(
\sign[1-(1+\gamma)U]\big)-\sigma =0.
\end{eqnarray*}
Similarly, $\Ex[\zeta\,|\,\xi=-\sigma]=0$.

To complete the proof of Remark~\ref{rem:2}, it remains to show that
part iii) of Assumption N is fulfilled. Indeed,
\begin{eqnarray*}
\frac{\log \Ex[e^{t\zeta}\,|\,\xi=\sigma]}{t^2\gamma \sigma^2}\eqserre\frac1{t^2\gamma\sigma^2}
\log\Bigg(e^{t\gamma\sigma}\frac{2+\gamma}{2(1+\gamma)}+e^{-t(2+\gamma)\sigma}\frac{\gamma}{2(1+\gamma)}\Bigg)\\
\eqserre \frac1{t^2\gamma\sigma^2}
\Bigg[t\gamma\sigma+\log\Bigg(1+\Bigg\{e^{-2t(1+\gamma)\sigma}-1\Bigg\}\frac{\gamma}{2(1+\gamma)}\Bigg)\Bigg].
\end{eqnarray*}
Applying the inequality of \cite[Lemma 3]{DT08} with
$\alpha_0=2(1+\gamma)/\gamma$ and $x=t\gamma \sigma$, we get
\begin{eqnarray*}
\frac{\log \Ex[e^{t\zeta}\,|\,\xi=\sigma]}{t^2\gamma \sigma^2}\leqserre
\frac1{t^2\gamma\sigma^2}
(t\gamma\sigma)^2\frac{(1+\gamma)}{\gamma}=1+\gamma
\end{eqnarray*}
and the desired result follows.

\subsubsection{Proof of Remark~\ref{rem:6}}
We start by computing the conditional moment generating function
(Laplace transform) of $\zeta$ given $\xi$:
\begin{eqnarray}
\Ex\big[e^{t\zeta}\,\big|\,\xi=a\big]\eqserre e^{-ta} \Ex\big[e^{t(\zeta+\xi)}\,\big|\,\xi=a\big]\nonumber\\
\eqserre e^{-ta} \Bigg(e^{t(1+\gamma)|a|}\Pb\big(\sign(a)>(1+\gamma)U\big)+e^{-t(1+\gamma)|a|}\Pb\big(\sign(a)<(1+\gamma)U\big)\Bigg)\nonumber\\
\eqserre e^{-ta}\Bigg(e^{t(1+\gamma)a}
\frac{2+\gamma}{2+2\gamma}+e^{-t(1+\gamma)a}\frac{\gamma}{2+2\gamma}\Bigg).\label{MomGen}
\end{eqnarray}
Using (\ref{MomGen}) we obtain
\begin{eqnarray*}
\Ex\big[e^{t(\zeta+\xi)}\big]\eqserre
\Ex\Big[\Ex\big[e^{t(\zeta+\xi)}\,\big|\,\xi\big]\Big]=
\frac{2+\gamma}{2+2\gamma}\Ex\big[e^{-t(1+\gamma)\xi}\big]+
\frac{\gamma}{2+2\gamma}\Ex\big[e^{t(1+\gamma)\xi}\big]=\Ex\big[e^{t(1+\gamma)\xi}\big],
\end{eqnarray*}
since the symmetry of $\xi$ implies that
$\Ex\big[e^{-t(1+\gamma)\xi}\big]=\Ex\big[e^{t(1+\gamma)\xi}\big]$
for every $t$. Thus, $\zeta+\xi$ has the same distribution as
$(1+\gamma)\xi$.

On the other hand, taking the derivatives of both sides of
(\ref{MomGen}) and using the fact that $\Ex[\zeta\,|\,\xi=a]$ equals
to the derivative of the moment generating function
$\Ex[e^{t\zeta}\,|\,\xi=a]$ at $t=0$, we obtain that
$\Ex[\zeta\,|\,\xi=a]=0$ for every $a\in [-B,B]$. To complete the
proof of Remark~\ref{rem:6} we apply \cite[Lemma 3]{DT08} to the
right hand side of (\ref{MomGen}). This yields
\begin{eqnarray*}
\log\Big(\Ex\big[e^{t\zeta}\,\big|\,\xi=a\big]\Big)\leqserre (t\gamma a)^2\frac{1+\gamma}{\gamma}\le (t B)^2\gamma (1+\gamma).
\end{eqnarray*}
Therefore, part iii) of Assumption N is satisfied with $v(a)\le
B^2$. This completes the proof of Remark~\ref{rem:6}.
%%--------------------------------------------------------------------}
%\bibliographystyle{alpha}
\section*{References}
\bibliographystyle{elsarticle-harv}
\bibliography{DT_JCSS_10}
%%--------------------------------------------------------------------}

\end{document}